\documentclass[english]{JHEP3}
\usepackage[T1]{fontenc}
\usepackage[latin9]{inputenc}
\usepackage{amsmath}
\usepackage{amssymb}
\usepackage{graphicx}
\usepackage{esint}

\newcommand\muR{\mu_{\rm\scriptscriptstyle R}}
\newcommand\muF{\mu_{\rm\scriptscriptstyle F}}
\newcommand\kT{k_{\rm\scriptscriptstyle T}}
\newcommand\KR{K_{\rm\scriptscriptstyle R}}
\newcommand\KF{K_{\rm\scriptscriptstyle F}}
\newcommand\nF{n_{\rm f}}
\newcommand\CF{C_{\rm F}}
\newcommand\CA{C_{\rm A}}
\newcommand\pT{p_{\rm\scriptscriptstyle T}}
\newcommand\pTH{p_{\rm\scriptscriptstyle T}^{\scriptscriptstyle H}}
\newcommand\pTZ{p_{\rm\scriptscriptstyle T}^{\scriptscriptstyle Z}}
\newcommand\alphaS{\alpha_{\rm\scriptscriptstyle S}}
\newcommand\LambdaMSB{\Lambda_{\rm\scriptscriptstyle \overline{MS}}}
\newcommand\CTEQsixM{{\tt CTEQ6M}}
\newcommand\MSTWNLO{{\tt MSTW2008NLO}}
\newcommand\POWHEG{{\tt POWHEG}}
\newcommand\PYTHIA{{\tt PYTHIA}}
\newcommand\POWHEGBOX{{\tt POWHEG BOX}}
\newcommand\MCFM{{\tt MCFM}}
\newcommand\Madgraphfour{{\tt MadGraph4}}
\newcommand\MCatNLO{{\tt MC@NLO}}
\newcommand\HJ{{\tt HJ}}
\newcommand\Z{{\tt Z}}
\newcommand\ZJ{{\tt ZJ}}
\newcommand\HJJ{{\tt HJJ}}
\newcommand\ZJJ{{\tt ZJJ}}
\newcommand\ggH{{\tt ggH}}
\newcommand\muQ{\mu_{\scriptscriptstyle Q}}
\newcommand\MH{M_{\scriptscriptstyle H}}
\newcommand\HThat{\hat{H}_{\scriptscriptstyle T}}
\newcommand\mH{M_{\scriptscriptstyle H}}
\newcommand\MZ{M_{\scriptscriptstyle Z}}
\newcommand\mZ{M_{\scriptscriptstyle Z}}
\newcommand\MINLO{{\tt MINLO}}
\newcommand\GH{\Gamma_{\scriptscriptstyle H}}
\newcommand\GZ{\Gamma_{\scriptscriptstyle Z}}

\makeatletter
\usepackage{cite}\usepackage[mathscr]{euscript}\preprint{CERN-PH-TH/2012-166\\ OUTP-12-11P\\MCnet-12-07}
\title{{\textbf{MINLO: Multi-scale improved NLO}}}

\author{Keith Hamilton\thanks{On leave from University College London.}\\
  Theory Division, CERN, CH--1211, Geneva 23, Switzerland\\
  E-mail: \email{keith.hamilton@cern.ch}
}
\author{Paolo Nason\\
  Theory Division, CERN, CH--1211, Geneva 23, Switzerland\\
  and INFN, sezione di Milano Bicocca\\
  E-mail: \email{paolo.nason@mib.infn.it}
}
\author{Giulia Zanderighi\\
  Rudolf Peierls Centre for Theoretical Physics, 1 Keble Road, University of Oxford, UK\\
  E-mail: \email{g.zanderighi1@physics.ox.ac.uk}
}

\abstract{In the present work we consider the assignment of the
  factorization and renormalization scales in hadron collider
  processes with associated jet production, at next-to-leading order
  (NLO) in perturbation theory. We propose a simple, definite
  prescription to this end, including Sudakov form factors to
  consistently account for the distinct kinematic scales occuring in
  such collisions. The scheme yields results that are accurate at NLO
  and, for a large class of observables, it resums to all orders the
  large logarithms that arise from kinematic configurations involving
  disparate scales. In practical terms the method is most simply
  understood as an NLO extension of the matrix element reweighting
  procedure employed in tree level matrix element-parton shower
  merging algorithms.  By way of a proof-of-concept, we apply the
  method to Higgs and Z boson production in association with up to two
  jets.}

\keywords{QCD, Phenomenological Models, Hadronic Colliders}

\usepackage{babel}

\makeatother

\usepackage{babel}
\begin{document}

\section{Introduction}
Perturbative QCD calculations depend on unphysical renormalization and
factorization scales.  Theoretical uncertainties are usually estimated
by varying the scales by a factor of two above and below their central
value.  In practice, in next-to-leading order (NLO) calculations the
central scale is often determined a posteriori by requiring either the
NLO corrections to be small, or the scale sensitivity to be
minimised. These stability criteria are usually motivated on the basis
that `bad' scale choices will give rise to large logarithms of the
ratio of the renormalization and/or factorization scale with respect
to scales characteristic of the process of interest and, hence, to
sizable corrections. These large corrections will induce in turn a
stronger scale dependence in the cross sections.

While it is certainly the case that unrepresentative scale choices can
lead to poor convergence of fixed order predictions, renormalization
and factorization scale logarithms are only one possible cause of
large higher order contributions. Significant physical contributions
may also arise from a number of other sources such as Sudakov effects,
large color factors, large $\pi^{2}$ terms, and the opening up of new
channels. Moreover, at least the dominant components of these
corrections take the form of double logarithmic contributions and
hence cannot be absorbed by any judicious choice of renormalization
and factorization scales.

As an illustrative example consider the familiar case of vector boson
production in hadronic collisions. For fully inclusive observables
there is no room for ambiguity in the choice of factorization scale,
which is naturally set equal to the mass of the vector boson, since
this scale limits the QCD radiation that accompanies the production
process. On the other hand, turning to the slightly more subtle case
of vector boson production in association with a jet, we are faced
with two scales: the vector boson mass and the jet transverse
momentum.  It is well known that this cross section carries (Sudakov)
double logarithms of the ratio of these two scales and that such terms
are large in the low transverse momentum region. By choosing the
renormalization and factorization scales on the basis that radiative
corrections be small, one is led to compensate such genuine physical
effects with unphysical scale logarithms.  Even if one takes the
extreme view that the renormalization and factorization scales are
arbitrary parameters that one is free `to tune' in making predictions,
it is difficult to see how the associated uncertainty could be
considered reliable or unbiased in the presence of theoretically
spurious compensation mechanisms.
 
In short, the stability criterion is potentially misleading, since it
attributes all large NLO corrections to these scale
logarithms. Both out of theoretical
correctness and pragmatism we are therefore motivated to look for an
unbiased method to choose the central scales, based on the kinematics
and dynamics of the process under study, rather than an a posteriori
observation of stability.

In the context of leading order matrix element-parton shower merging
algorithms
\cite{Catani:2001cc,Mangano:2001xp,Lonnblad:2001iq,Krauss:2002up,Mrenna:2003if,Mangano:2004lu},
an unbiased method for assigning the factorization and renormalization
scales is essential and put to good effect. In practice, the kinematic
configuration of the process is associated with the most probable
branching history by an exclusive jet clustering algorithm. The
transverse momentum at each branching defines the renormalization
scale for the corresponding factor of $\alpha_{\mathrm{S}}$ at the
vertex, while the factorization scale is associated with the matrix element - parton shower merging
scale. Furthermore, a recipe is given for including Sudakov form
factors, accounting for the large double logarithms that arise when
the clustered event contains well separated scales. The net effect of
the scale assignment and Sudakov factors is to incorporate,
consistently, all large logarithms, i.\ e.\ renormalization,
factorization and Sudakov logarithms, associated with rendering the
event exclusive with respect to radiation above the matrix
element-parton shower merging scale --- the scale beneath which the
parton shower is used to populate the remaining phase space.%
\footnote{This is not the case for small $x$ or threshold logarithms,
  that we are not considering in this context.}

It therefore seems appropriate to adapt the calculation of NLO cross
sections such that the Born term is evaluated with the scales and
Sudakov form factors prescribed by the CKKW method
\cite{Catani:2001cc,Krauss:2002up,Mrenna:2003if}. In the present work
we pursue this possibility and construct such a procedure in
accordance with the following generic requirements:
\begin{itemize}
\item the full result has formal NLO accuracy, therefore the scale
  variation around the central values is formally of
  next-to-next-to-leading (NNLO) order;
\item the accuracy and the smooth behaviour near the Sudakov regions
  is comparable to that of the corresponding tree-level calculation in
  the adopted CKKW scheme;
\item the procedure is simple and easily implemented for any NLO
  parton level generator, requiring only minor work on top of the NLO
  calculation available.
\end{itemize}

The procedure we propose is based upon two simple observations. The
first one concerns the choice of the renormalization scale $\mu_R$.
To this end let us note that NLO cross sections have the formal
structure
\begin{equation}
\frac{\mathrm{d}\sigma}{\mathrm{d}\Phi}=\alpha_{\mathrm{S}}^{N}(\muR)\, B+\alpha_{\mathrm{S}}^{N+1}(\muR)\left[V+N\, b_{0}\,\log\frac{\muR^{2}}{Q^{2}}\, B\right]+\alpha_{\mathrm{S}}^{N+1}(\muR)\, R\,,\label{eq:NLOfixed}
\end{equation}
where $B$ denotes the Born term and $R$ the real corrections to it.
The virtual corrections are shown in parenthesis, their explicit
renormalization scale dependence being proportional to the Born term,
where $b_{0}$ is the one loop beta function coefficient
\begin{equation}
b_{0}=\frac{33-2\,\nF}{12\,\pi}\,,
\end{equation}
and $Q$ is a momentum scale representative of the leading order
kinematics. The explicit $\muR$ dependence of the virtual corrections
is such that the variation of eq.~(\ref{eq:NLOfixed}) with respect to
changing the renormalization scale is of order
$\alpha_{\mathrm{S}}^{N+2}$; terms of order $\alpha_{S}^{N+1}$ induced
by varying $\muR$ in the Born and virtual contributions cancel exactly
due to the renormalization group equation.

From here it is clear that should we choose to evaluate the $N$
coupling constants in the Born term at different scales $\{\mu_{i}\}$,
as in the matrix-element-parton shower merging algorithms, in order
for NLO scale compensation to take place eq.~(\ref{eq:NLOfixed}) must
generalise to
\begin{equation}
\frac{\mathrm{d}\sigma}{\mathrm{d}\Phi}=\prod_{i=1}^{N}\alphaS(\mu_{i})\, B+\alphaS^{N+1}(\muR')\left[V+b_{0}\sum_{i=1}^{N}\log\frac{\mu_{i}^{2}}{Q^{2}}\, B\right]+\alphaS^{N+1}(\muR'')R\hspace{0.25em},\label{eq:skkwscales}
\end{equation}
where the scales $\muR'$ and $\muR''$ in the virtual and real terms
are irrelevant from the point of view of scale compensation:
$\alphaS(\muR)-\alphaS(\muR')\approx\mathcal{O}(\alphaS^{2})$.  In
eq.~(\ref{eq:skkwscales}), scale compensation takes place
independently for each of the $\mu_{i}$ that is varied. While it may
be a relatively straightforward task to evaluate $N$ coupling
constants at $N$ scales for the Born term, virtual corrections in NLO
calculations are usually expressed in terms of a single
renormalization scale only. However, by simply setting $\muR$ in the
virtual term to be the geometric mean of the $\mu_{i}$ in
eq.~(\ref{eq:skkwscales})
\begin{equation}
\muR=\left(\prod_{i=1}^{N}\mu_{i}\right)^{\frac{1}{N}}\hspace{0.25em}\,,
\end{equation}
and evaluating the $N$ coupling constants in the Born term at scales
$\mu_{i}$, we arrive at an expression precisely of the form in
eq.~(\ref{eq:skkwscales}).
Equivalently, we can evaluate the virtual term at some fixed
scale $\mu_0$ and explicitly add the following contribution
\begin{equation}
\alphaS^{N+1}(\muR')\times b_0\sum_{i=1}^N \log\frac{\mu_i^2}{\mu_0^2}B\;.
\end{equation}

The second ingredient that is needed in order to maintain NLO accuracy
has to do with the Sudakov form factors that are included in the Born
term in the CKKW approach. These form factors, when expanded in powers
of $\alphaS$, lead to terms of order $\alphaS^{N+1}$, i.e. of the NLO
level of accuracy. These terms should be subtracted in order to
maintain NLO accuracy.

The choice of scales in the arguments of each power of $\alphaS$ in
the real and the virtual terms, the exact definition of the
subtraction term arising from the expansion of the Sudakov form
factors, and the inclusion of the Sudakov form factors in the real and
virtual terms, remain to a large extent arbitrary as far as the NLO
accuracy is concerned. We will however further constrain these
choices, in such a way that the virtues of the CKKW result at leading
order are maintained once radiative corrections are included. We defer
the discussion of these and further details to the main body of the
article.

The method presented in this paper can be applied in order to improve
the prediction for inclusive quantities in any NLO calculation. It is
however particularly advantageous in the context of interfacing NLO
calculations to parton shower
programs~\cite{Frixione:2002ik,Nason:2004rx}. In the \POWHEG{}
framework\cite{Nason:2004rx,Frixione:2007vw}, for example, the
underlying Born structure of the event is generated with a probability
proportional to the NLO inclusive cross section at a given point in
the Born phase space. This cross section can be evaluated
using the prescription advocated in the present work, leading to a
considerable improvement in reliability near the Sudakov regions.

The paper is organized as follows. In Section
\ref{sec:Theoretical-considerations} we review briefly the CKKW method
for matrix element-parton shower merging. In Section \ref{sec:method}
we present in detail our prescription.  A theoretical discussion
regarding the interplay of the scale choices and Sudakov form factors
is given in Section~\ref{sec:further}. In
Section~\ref{sec:phenomenology}, as an example, we apply our method to
the case of Higgs and $Z$ production in association with one or two
jets. Finally, we present our conclusions in Section \ref{sec:conc}.
In the Appendix we give the exact expression of the Sudakov form
factors at next-to-leading logarithmic (NLL) accuracy that we used to
obtain the results presented here.

\section{Summary of the CKKW formalism\label{sec:Theoretical-considerations}}

We first briefly summarize the standard CKKW
procedure~\cite{Catani:2001cc,Krauss:2002up,Mrenna:2003if}. We
consider a production process in hadronic collisions. The CKKW
formalism requires that we recursively cluster the coloured partons in
the event using a $\kT$-clustering
algorithm~\cite{Ellis:1993tq,Catani:1993hr}, in order to reconstruct
the most likely branching history. The $\kT$-clustering should be
consistent with the flavour structure, i.e. a pair of partons can only
be clustered if it can come from a single parton, and the appropriate
flavour is assigned to the parton arising from the merging. At each of
the vertices $i$ ($i=1,\ldots,n$) of the branching history, one
assigns a nodal scale $q_{i}$, equal to the relative transverse
momentum value at which the clustering has taken place. In the CKKW
formalism one also assigns a resolution scale $Q_{0}$, meaning that
the cross section is interpreted as being inclusive for all radiation
below $Q_{0}$.

The recursive procedure ends when no further clustering is possible
and we refer to the remaining ensemble of particles as the
\emph{primary system}.%
\footnote{In processes like $W+\mathrm{jets}$ production, the
  clustering typically stops when all jets are clustered away. In the
  case of jet production, clustering should stop when at least two
  jets are left.}  We assign it a scale equal to its invariant mass
$Q$. The CKKW cross section is obtained by taking the tree-level
matrix element, with the strong couplings associated with each node
evaluated at the corresponding scale. The remaining $m=N-n$ powers of
the strong coupling%
\footnote{In the case of Higgs production in gluon fusion, for
  example, there will be always at least two powers of $\alphaS$
  associated with the primary system.}  are associated with the
primary system, and are evaluated at the scale $Q$. Intermediate lines
between nodes $i$ and $j$ in the branching history are furthermore
assigned a Sudakov form factor
\begin{equation}
\frac{\Delta_{f_{ij}}(Q_{0},q_{i})}{\Delta_{f_{ij}}(Q_{0},q_{j})},
\label{eq:Sudff}
\end{equation}
where $f_{ij}$ is the flavour of the line joining $i$ and $j$, where
$i$ is the node closest to the primary vertex ($q_i> q_j$).  External
lines have Sudakov form factors equal to $\Delta_{f}(Q_{0},q_{i})$,
where $i$ is the node connected to the external line.

The general form of the Sudakov exponent is
\begin{equation}
\Delta_{f}(Q_{0},Q)=\exp\left[-\int_{Q_{0}}^{Q}dq\,\frac{2C_{f}}{\pi}\frac{\alphaS(q)}{q}\left(\log\frac{Q}{q}-B_{f}\right)\right],\qquad
f = q,g\,,
\end{equation}
where $C_{g}=C_{A}$, $B_{g}=\pi b_{0}/C_{A}$ or $C_{q}=C_{F}$,
$B_{q}=3/4$ for gluon or quark lines respectively. Using the leading
logarithmic expression for $\alphaS$, we can compute the Sudakov form
factor analytically. We obtain
\begin{equation}
\Delta_{f}(Q_{0},Q)=\exp\left[-\frac{C_{f}}{\pi b_{0}}\left\{ \log\frac{\log\frac{Q^{2}}{\Lambda^{2}}}{\log\frac{Q_{0}^{2}}{\Lambda^{2}}}\left(\frac{1}{2}\log\frac{Q^{2}}{\Lambda^{2}}-B_{f}\right)-\frac{1}{2}\log\frac{Q^{2}}{Q_{0}^{2}}\right\} \right].\label{eq:deltafLO}
\end{equation}
A more detailed analysis, adequate for NLL accuracy, is presented
in appendix~\ref{app:sudak}.

Expanding eq.~(\ref{eq:deltafLO}) in powers of $\alphaS$ we get
\begin{eqnarray}
\Delta_{f}(Q_{0},Q)&=&1+\Delta_{f}^{(1)}(Q_{0},Q)+\mathcal{O}(\alphaS^{2}),
\\
\Delta_{f}^{(1)}(Q_{0},Q)&=&-\frac{C_{f}}{\pi}\alphaS
\left[\frac{1}{4}\log^{2}\frac{Q^{2}}{Q_{0}^{2}}
-\log\frac{Q^{2}}{Q_{0}^{2}}B_{f}\right],\label{eq:SUDexpanded}
\end{eqnarray}
that represents the effective NLO correction that is already included
in the Born term when we use the CKKW prescription, and will
eventually be subtracted in our method.

Finally, we note that in the CKKW algorithm the factorization scale
in the parton density functions is set to $Q_{0}$, the
matrix element-parton shower merging scale.
Each event from the tree-level matrix element generator, when
reweighted to include these Sudakov form factor and scale settings,
is then passed to a parton shower simulation, constrained in such
a way that no further radiation is generated at scales above $Q_{0}$.
Hence, the distribution of radiation resolved at scales above $Q_{0}$
is governed by exact tree-level matrix elements with the remaining
phase space filled by the parton shower.

In the CKKW scheme, inclusive configurations with the maximum number
of partons in the matrix elements are treated differently~\cite{Mrenna:2003if}.
In this case, the scale $Q_{0}$ is taken
equal to the lowest merging scale. Hence, when interfacing the
reweighted tree level events to the parton shower, all higher jet
multiplicities, for which no tree-level matrix element is available,
will be consistently generated by the shower.

In the context of our approach, the natural choice of $Q_0$ is the
same one adopted in the CKKW scheme for the highest multiplicity
sample. More specifically, in our case $Q_{0}$ should be set equal to
the scale of the first merging for Born level kinematics (i.e. the
Born and the virtual), and the scale of the second merging for the
real kinematics. This is easily understood with the following
example. When we consider $Z+\mathrm{jet}$ production, with the jet
transverse momentum equal to $p_{T}$, we clearly imply that the jet we
are considering is the hardest one and thus that its $p_{T}$ limits
the scale of all other jets. In the case of the real emission in an
NLO calculation, the lowest merging scale corresponds to integrating
over further radiation inclusively, hence, its merging scale (the
first one in this case) is skipped.

\section{Formulation of the method}\label{sec:method}

We now formulate our complete prescription, \MINLO{} for Multi-scale
improved NLO, including the choice of scales appearing in the coupling
constants associated with the NLO corrections, the
inclusion of the Sudakov form factors in the virtual and real
contributions, and how to perform the subtraction of the term in
eq.~(\ref{eq:SUDexpanded}). We recall that $Q$ is the scale of the primary configuration, $q_{1}\ldots q_{n}$ are
the remaining clustering scales in increasing order, and that
generally we have $m$
powers of $\alphaS$ (where $m$ can be zero) associated with the
primary process. In the case of the real cross section, there will be
also a smallest clustering scale $q_{0}$, corresponding to the first
clustering. As discussed above, we will always fix the scale $Q_{0}$ entering eq.~(\ref{eq:Sudff}) to $q_{1}$. We then proceed as follows:
\renewcommand{\labelenumi}{\roman{enumi}.}
\begin{enumerate}
\item We perform the $\kT$ clustering of the event, determine the
  scales $Q$, $q_1\ldots q_n$, and eventually $q_0$ for the real term,
  and construct the event skeleton.
  We cluster only partons that are
  compatible in flavours, i.e. gluons with gluons, yielding gluon
  pseudopartons, gluons and quarks, yielding quark pseudopartons with
  the same flavour, and quarks with antiquarks of opposite flavour,
  leading to gluon pseudopartons. We set $Q_0=q_1$.

  It may occur that the scale of the primary process $Q$ turns out to
  be smaller than the last clustering scale. This happens, for example
  in the production of a massive boson recoiling against a hard jet,
  with transverse momentum larger than the boson mass. In these cases
  we will take $Q=q_n$. Notice that this choice is not fully motivated by the CKKW
  approach, which instead deals with naturally ordered radiation.
  Although this case is interesting on its own, being perhaps related
  to the giant $K$-factor issues~\cite{Rubin:2010xp}, we will not
  pursue it further in the present work.

\item $n$ powers of the coupling constant in the Born, virtual
  and real contributions will be evaluated at the
  scales $\mu_{1}\ldots\mu_{n}$, with $\mu_i= \KR\, q_{i}$ ($i=1\ldots
  n)$ (the value of $\alphaS$ to be used in the real and
  virtual contributions for the
  $(n+m+1)^{\rm th}$ power of the coupling constant
  is specified at point VI).
  $\KR$ is the renormalization scale factor, equal to 1 for the
  central value, and typically varied between $0.5$ and $2$ in order
  to study scale variation uncertainties. The $m$ strong coupling
  constants associated with the primary system will be taken equal to
  $\KR Q$. 

\item The renormalization scale explicitly appearing in the virtual
  corrections is set to
  $\muR=\left((\muQ)^{m}\times\prod_{i=1}^{n}\mu_{i}\right)^{\frac{1}{m+n}}$,while
  the factorization scale $\muF$, appearing explicitly in the
  collinear subtraction remnants and in all parton densities functions
  (pdf's), is assigned the scale $\KF q_1$, where $\KF$ is the
  factorization scale factor.

\item The Sudakov form factors for all the skeleton lines will be
  included for the Born, virtual and real contributions. For the
  latter, as already remarked above, we include the Sudakov form
  factors corresponding to the branching history obtained after the
  first clustering. Notice that the external lines that join at the
  first node have $\Delta(Q_{0},q_{1})=1$, since $Q_0=q_1$.

\item The subtraction of the NLO contribution present in the CKKW Born
  term is performed by replacing
\begin{equation}
  B\Rightarrow B\times\left(1-\sum_{ij}\left[\Delta_{f_{ij}}^{(1)}(Q_{0},q_{i})-\Delta_{f_{ij}}^{(1)}(Q_{0},q_{j})\right]-\sum_{l}\Delta_{f_{l}}^{(1)}(Q_{0},q_{k_{l}})\right),\label{eq:subtrfac}
\end{equation}
where the first sum extends over all pairs of nodes $i,j$, with
$q_{i}>q_{j}$, connected by a line of flavour $f_{ij}$, and the second
one runs over all external lines $l$ connected to nodes $k_{l}$ 
(excluding $k_{l}=1$, which vanishes).
\item For the value of $\alphaS$ to be used in the $(n+m+1)^{\mathrm{th}}$
power of $\alphaS$ appearing in the real and virtual cross section,
and also appearing in eq.~(\ref{eq:subtrfac}), we propose 
\begin{equation}\label{eq:alphanpone}
\alphaS^{(n+m+1)}=\frac{1}{n+m}\left(\sum_{i=1}^{n}\alphaS(\mu_{i})+m\alphaS(\muQ)\right).
\end{equation}
The logic for this choice is the following. Large QCD corrections can
be viewed as being associated with the nodal scales in the branching
history, and can thus be viewed as an $\alphaS$ factor evaluated at
the nodal scales times the Born cross section, one for each node.  The
sum of them will lead to a sum of $\alphaS$ values taken at each nodal
scale.  As far as the subtraction term in eq.~(\ref{eq:subtrfac}) is
concerned, here we make the same choice performed in the NLO terms,
since the subtraction term is meant to subtract large corrections
arising in the NLO terms and already resummed when the full Sudakov
form factors are multiplied by the Born term.

Notice that in eq.~(\ref{eq:subtrfac}) we could have instead used the
same value of $\alphaS$ that appears in the Sudakov form factor,
rather than the one given in eq.~(\ref{eq:alphanpone}).  By sticking
to the present choice, we may be artificially reducing the scale
dependence of the whole result. The exploration of this alternative,
as well as many other possible variations on the method, will be left
to future work. The purpose of the present work is just to present the
essential features of the method, and thus we will stick to a definite
choice among all possible options.
\end{enumerate}
To further motivate the above prescription, we make the following
remarks. First of all, the inclusion of Sudakov form factors and
running couplings in the NLO corrections, with essentially the same
prescription as in the Born term, guarantees that also when NLO
corrections are included, we recover in the Sudakov regions the same
smooth behaviour that was present in the Born term alone thanks to the
CKKW procedure.  A second important remark has to do with the form of
the subtraction term arising from eq.~(\ref{eq:subtrfac}). We notice
that this term has precisely the same couplings and Sudakov form
factors present in the NLO term. It is thus constructed in such a way
as to have an optimal cancellation of the large Sudakov logs arising
in the NLO corrections, that are already present in exponentiated form
in the Born Sudakov form factor.

\section{Interplay between scale choices and Sudakov form factors}
\label{sec:further}

It is often easy to find conflicting motivations for the choice of
scale in an NLO calculation. Consider the example of Higgs plus jet
production, assuming that the jet momentum is substantially lower than
the Higgs mass. This process is of order $\alphaS^{3}$ at the Born
level and one may be inclined to believe that one out of the three
powers of $\alphaS$, being associated with the radiated jet, should be
taken of the order of the jet transverse momentum, while the other two
should be of the order of the Higgs mass, and that the factorization
scale should be an intermediate scale between the two. On the other
hand, if we recall that our cross section describes the hardest jet,
and should be viewed as inclusive in all radiation with clustering
scale below the $p_{T}$ of the jet, we would reach the conclusion that
the factorization and renormalization scales for all powers of
$\alphaS$ should be taken equal to the jet transverse momentum. This
is because all gluon propagators and external lines (including the
incoming ones) are limited in virtuality by the jet transverse
momentum (the internal line by kinematics and the external lines
because radiation with merging scale above $p_{T}$ is not allowed).

This apparent conflict illustrates how failing to consider the effect
of Sudakov form factors when dealing with the choice of the scales can
lead to inconsistent conclusions. First of all, we should recall that
Sudakov double logarithms are formally more important than
renormalization or factorization scale logarithms, since the latter
lead only to single logs. Furthermore, it should also be remembered that
some sub-leading terms in the Sudakov logarithms are precisely there
to compensate the mismatch between different scales at connected
vertices. The purpose of this section is to further elaborate upon
these points, and to demonstrate that a scale assignment, in the
framework of multi-jet processes, cannot be consistently discussed if
Sudakov form factors are not properly included.

\begin{figure}[tbh]
\begin{centering}
\includegraphics[scale=0.35]{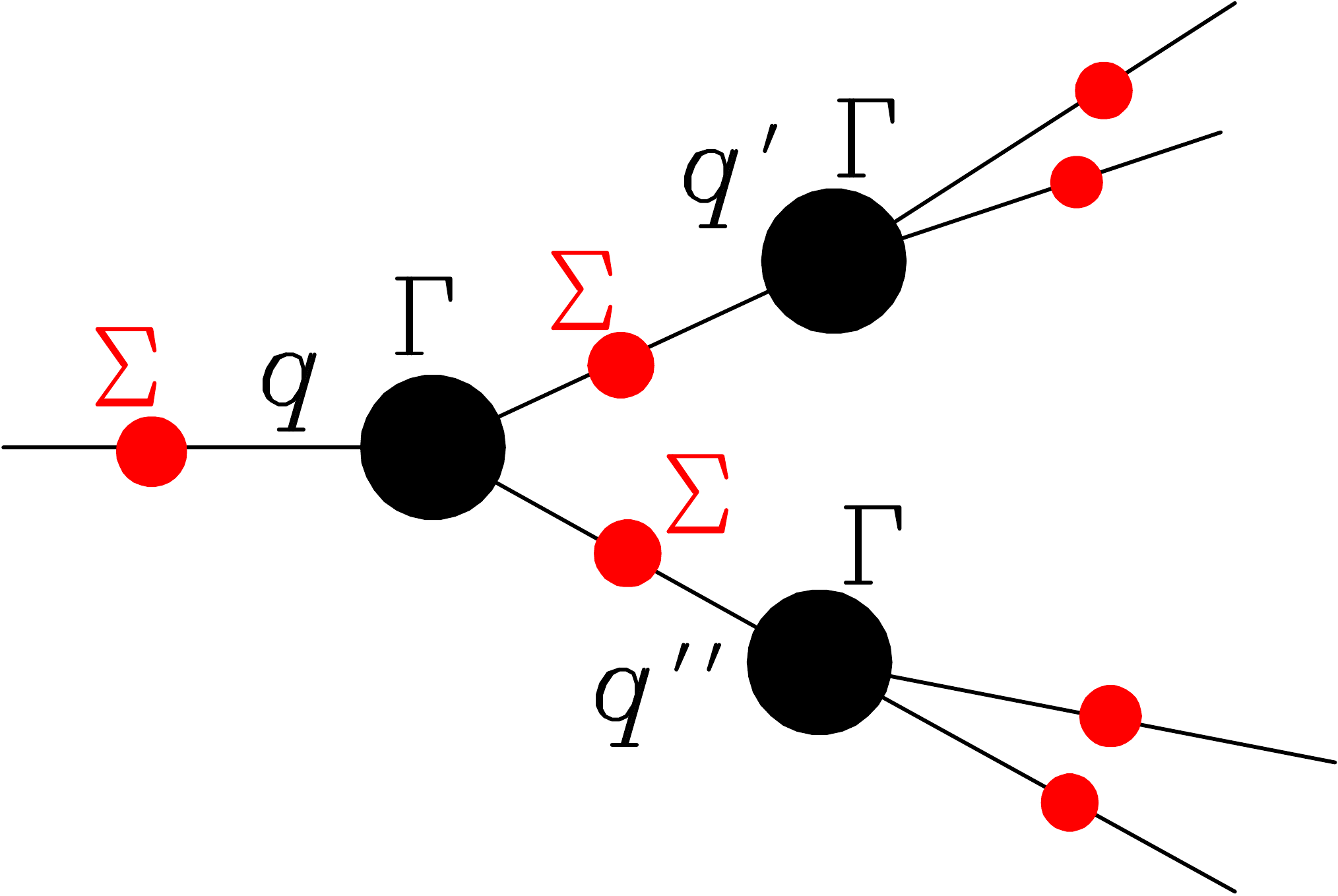} 
\par\end{centering}
\caption{Schematic leading logarithmic corrections to a branching history.\label{fig:virtLL}}
\end{figure}

Consider the simple example of quantum field theories without
infrared divergences, like Yukawa theories or $\Phi^{3}$ in 6
space-time dimensions. In such theories, we may look for the dominant
virtual corrections to a branching process by including leading
logarithmic virtual corrections to all vertices and internal lines, as
illustrated in fig. \ref{fig:virtLL},
where $\Gamma$ represent the one particle irreducible vertex
corrections, and $\Sigma$ represent self energies. In these theories
(and also in gauge theories in physical gauges), the vertex
corrections are infrared finite when at least one external line is off
shell, so that they are dominated by the largest virtuality. Thus, in
fig.~\ref{fig:virtLL}, for the leftmost vertex we have
\begin{equation}
\Gamma(q,q',q'')\approx\Gamma(q,q,q),
\end{equation}
and we will write for simplicity
\begin{equation}
\Gamma(q,q,q)=\Gamma(q).
\end{equation}
On the other hand, we have
\begin{equation}
\Gamma(q)[\Sigma(q)]^{\frac{3}{2}}=\frac{g(q)}{g},
\end{equation}
where $g(q)$ is the running coupling at the scale $q$. Thus, if at
each vertex with an incoming line having virtuality $q$ we substitute
\begin{equation}
\Gamma(q)=\frac{g(q)}{g [\Sigma(q)] ^{\frac{3}{2}}},
\end{equation}
we immediately see that the net effect of the insertion of vertex and
self energy corrections is the inclusion of the running coupling
constant at the scale of the incoming virtuality for each vertex, and
of a factor
\begin{equation}
\sqrt{\frac{\Sigma(q')}{\Sigma(q)}}
\end{equation}
for each line. This yields in the cross sections, i.e. in the
full squared amplitude, a factor
\begin{equation}
\Delta(q',q)=\frac{\Sigma(q')}{\Sigma(q)},
\end{equation}
which is the Sudakov form factor.

It is interesting to look in detail to what happens in the case of
Higgs plus jet production and how the apparent contradictions arising
from naive scale assignments are solved if the full Sudakov form
factors are included. The couplings and Sudakov factors that multiply
the tree level amplitude in the CKKW approach yield the factor
\begin{equation}\label{eq:sudexample}
F =  \alphaS^{2}(\MH)\alphaS(p_{T})\left\{ \exp\left[-\frac{C_{A}}{\pi
      b_{0}}\left\{
    \log\frac{\log\frac{Q^{2}}{\Lambda^{2}}}
     {\log\frac{Q_{0}^{2}}{\Lambda^{2}}}
       \left(\frac{1}{2}\log\frac{Q^{2}}{\Lambda^{2}}-\frac{\pi
      b_{0}}{C_{A}}\right)-\frac{1}{2}\log\frac{Q^{2}}{Q_{0}^{2}}\right\}
    \right] \right\}^2\,.
\end{equation}
Notice that we
have only the two powers of the Sudakov form factors, associated with
the incoming internal and external lines that join at the Higgs
production vertex. The remaining two lines join at the first node, hence their Sudakov form factor is one, since $Q_{0}^{2}=p_{T}^{2}$.
Eq.~(\ref{eq:sudexample}) in turn leads to
\begin{eqnarray}
F &=&   \alphaS^{2}(\MH)\alphaS(p_{T})
  \left(\frac{\log\frac{Q^{2}}{\Lambda^{2}}}{\log
   \frac{Q_{0}^{2}}{\Lambda^{2}}}\right)^2\exp\left[-\frac{C_{A}}{\pi
      b_{0}}\left\{
    \log\frac{\log\frac{Q^{2}}{\Lambda^{2}}}{\log
    \frac{Q_{0}^{2}}{\Lambda^{2}}}\log
    \frac{Q^{2}}{\Lambda^{2}}-\log\frac{Q^{2}}{Q_{0}^{2}}\right\}
    \right] \nonumber \\
 & \approx&\;\alphaS^{3}(p_{T})\exp\left[-\frac{C_{A}}{\pi
      b_{0}}\left\{
    \log\frac{\log\frac{Q^{2}}{\Lambda^{2}}}{\log
    \frac{Q_{0}^{2}}{\Lambda^{2}}}\log
    \frac{Q^{2}}{\Lambda^{2}}-\log\frac{Q^{2}}{Q_{0}^{2}}\right\}\right]\,,
\label{eq:scalespt}
\end{eqnarray}
where we have taken $Q_{0}^{2}=p_{T}^{2}$ and $Q=\MH$. 
Notice that
\begin{equation}
\exp\left[-\frac{C_{A}}{\pi b_{0}}\left\{ \log\frac{\log\frac{Q^{2}}{\Lambda^{2}}}{\log\frac{Q_{0}^{2}}{\Lambda^{2}}}\log\frac{Q^{2}}{\Lambda^{2}}-\log\frac{Q^{2}}{Q_{0}^{2}}\right\} \right]\approx1-\frac{C_{A}}{\pi}\alphaS\frac{1}{2}\log^{2}\frac{Q^{2}}{Q_{0}^{2}}+\mathcal{O}\left(\alphaS^{2}\right)\;,
\end{equation}
i.e. the pure Sudakov double logarithm.  Thus, applying the CKKW
prescription leads to the conclusion that the scale choice for
$\alphaS$ is $\pT$ for all powers of $\alphaS$, provided a pure LL
Sudakov form factor is included. If one assigns the scale $\pT$ to
one power of $\alphaS$, and $\MH$ to the remaining two, then the full
NLL Sudakov form factor should be included, that takes care of the
scale mismatch.

Thus, the intuitive argument of assigning the same $p_{T}$ scale to
all coupling constant is in a sense correct, but one should not forget
that double log Sudakov terms are formally more important than scale
logarithms.

\section{Phenomenology}\label{sec:phenomenology}
In order to test our prescription, we have implemented it in the
\POWHEGBOX{} \cite{Alioli:2010xd} in a fully generic way, so that it
can be applied to any process of interest. In this context, we have
performed a variation over the scheme presented in Section
\ref{sec:method}, regarding the first clustering in the real emission
contributions. Since the \POWHEGBOX{} already provides a first
clustering, corresponding to the mapping of the real emission
configuration to its underlying Born structure, we have relied on this
mapping rather than performing this clustering explicitly using the $\kT$
algorithm. This procedure is formally equivalent to the one given in
Section~\ref{sec:method}, and it has the advantage of greater
simplicity.
We have used $R=1$ in our $\kT$ clustering procedure.
The Sudakov form factors have been coded both with the
expressions of eq.~(\ref{eq:deltafLO}), and with the full NLL
dependence presented in the appendix. It turns out that the two
expressions differ very little if the value of $\Lambda$ used in the
leading order expression is taken equal to $\LambdaMSB$. However, we
have produced our results using for the Sudakov exponent the full
expression given in appendix~\ref{app:sudak}.

Our study will focus on two examples: Higgs production via gluon
fusion and $Z$ production, both in association with one or two
jets. The (infinite top mass limit) Higgs production NLO calculations
are taken from
refs.~\cite{deFlorian:1999zd,Ravindran:2002dc,Campbell:2006xx,Campbell:2010cz,Campbell:2012am}. They
will be referred to as the \HJ{} (for Higgs plus one jet) and \HJJ{}
(for Higgs plus two jets) in the present work.  The $Z+$~jet cross
section (\ZJ{} in this paper) is taken from ref.~\cite{Alioli:2010qp}.
A $Z+2$~jets \POWHEGBOX{} implementation has appeared in
ref.~\cite{Re:2012zi}. However, the relevant code is not fully public.
We thus implemented a new $Z+2$~jets (\ZJJ{}) \POWHEGBOX{} generator
using the automatic \Madgraphfour{} interface developed in
ref.~\cite{Campbell:2012am}, taking the virtual corrections from the
\MCFM{} package~\cite{Giele:1993dj,Bern:1997sc,Campbell:2002tg}.

We will refer to the results obtained with the method presented in
this work as \MINLO{} (for Multi-scale Improved NLO). All the
calculations are performed for the LHC at a centre of mass energy of 7
TeV. The Higgs mass is always taken equal to $120\;$GeV. We have used
the \CTEQsixM{} parton density functions 
\cite{Pumplin:2002vw} for Higgs production, 
\MSTWNLO~\cite{Martin:2009iq} for Z production processes, and the
$k_T$ algorithm for jets, with $R=0.5$, as implemented in
{\tt FastJet}~\cite{Cacciari:2011ma}. In the Higgs boson production case,
the full cross section is reported, with no branching ratios.  All results
concerning $Z$ production include the branching fraction for $Z\to e^+ e^-$.
A mass
window from $60\;{\rm GeV}$ to $\MZ+15\GZ$ was used for the $Z$
virtuality ($\GZ = 2.495$ GeV), while for the Higgs Boson virtuality
we have considered the window from $\MH-15\GH$ to $\MH+15\GH$
($\GH=5.75\cdot 10^{-3}$ GeV).

When showing showered \POWHEG{} results for comparison, 
these will always be generated interfacing \POWHEG{} with
\PYTHIA{} 6.4.25 \cite{Sjostrand:2006za}, using the Perugia-0
tune ({\tt PYTUNE(320)}), with hadronization and underlying event turned
off.

We will compare the \MINLO{} results also to
standard NLO results obtained with conventional scale
choices. In particular, a fixed scale choice,
labelled `FXD' in the figures, will correspond
to the scales central values equal to the mass
of the heavy boson in all cases. A running
scale (labelled `RUN') will also be considered.
It will be taken equal to the jet transverse
momentum in both the $H+1$~jet and $Z+1$~jet
processes, since this is the scale that one
would adopt following the intuitive reasoning
of Section~4. In the $H+2$~jets case, the
running scale will be taken equal to $\HThat$,
defined as
\begin{equation}
\HThat=\sqrt{{\mH}^2+{\pTH}^2}+\sum_i \pT^{(i)}\,, 
\end{equation}
where the sum runs over all partons in the event. 
In the $Z+2$~jets case, the running scale will
be taken equal to $\HThat/2$. The $\HThat$
scale is quite popular in multijet processes, and,
in particular, $\HThat/2$
seems to be the preferred scale
for $W$ and $Z$ production in association with
jets~\cite{Berger:2010zx}.

\subsection{Preliminary considerations}
Before discussing the goal of our study, it is useful to clarify what
we expect from our method by making a couple of consideration
regarding the CKKW algorithm when applied to the tree level cross
sections. Consider for example, Higgs production plus $n$ partons.
Because of the unitarity of the shower, in the parton shower
approximation of the Higgs plus $n$ partons process, by integrating
over the last splitting, one recovers exactly the shower approximation
to the cross section for Higgs plus $(n-1)$ partons.  One expects
something similar to happen for the CKKW formula.%
\footnote{Notice that, because of the presence of the Sudakov form
  factors, the CKKW formula is integrable in the full phase space,
  provided we avoid integrating over the Landau pole of the running
  coupling constant by, for example, freezing the coupling at a scale
  just above it.}  However, in the CKKW case unitarity is not exact,
and this feature is only approximate. In the simplest case of a single
radiated parton we can easily prove that by integrating out the
radiated parton we recover the Born cross section up to corrections of
order $\alphaS$. In the case of more complex configurations,
sub-leading logarithms can arise, and a sound conclusion is more
difficult to reach.

In order to assess the performance of a \MINLO{} result, we need to
compare it to other calculations that give a reasonably good
description of the Sudakov region. So, for example, we will compare
the \MINLO{} \HJ{} result to the showered, parton level \POWHEG{}
result for inclusive Higgs (\ggH{}). If we look, for instance, at the
Higgs transverse momentum distribution, the \POWHEG{} \ggH{} result
gives a correct description of the Sudakov region and, furthermore,
its integral yields the NLO accurate total Higgs production cross
section. On the other hand, it describes the tail of the Higgs
transverse momentum distribution only with LO accuracy.  By contrast,
the \MINLO{} result is instead NLO accurate at relatively large
transverse momenta and LO accurate for the integral of the whole
distribution.

In the case of heavy boson production in association with two jets,
like in Higgs plus dijet production, we will compare the \MINLO{}
\HJJ{} result with the \POWHEG{} \HJ{} one, enhanced with the
\MINLO{} prescription. Here
we expect the \POWHEG{} \HJ{} result to give a good description of the
Sudakov region associated with the emission of the second
parton. Integrating out the second parton emission in distributions
that are inclusive in the hardest jet, one achieves NLO accuracy. On
the other hand, only LO accuracy is achieved for the production of two
widely separated jets.  Conversely, the \MINLO{} \HJJ{} calculation
has full NLO accuracy for Higgs plus two jets and leads to LO accuracy
for Higgs plus one jet distributions.

We remark here that the standard NLO calculations are not integrable
over the full phase space. Thus, for example, the \HJJ{} standard NLO
result does not yield a finite cross section for Higgs plus one jet
distributions, while in the \MINLO{} approach a sensible result is
obtained (thanks to the damping of the Sudakov form factors), although only accurate at leading order.

The fact that the \MINLO{} NLO calculation is finite is a
remarkable advantage over the usual fixed order calculations, since,
for instance, when generation cuts are imposed in order to obtain
finite cross sections, one needs to make sure that the cuts are low
enough so that final results are not sensitive to them. However,
making generation cuts too low renders NLO calculations inefficient,
so that usually an appropriate, delicate compromise needs to be found.
Another feature that is worth stressing in the \MINLO{} result is the
improved stability of the inclusive distributions as the Sudakov
regions are reached. This is not only due to the Sudakov suppression
factor, but also to the fact that Sudakov logarithms arising at fixed
order in the NLO corrections are compensated by the inclusion of the
subtraction terms of eq.~(\ref{eq:subtrfac}), which have exactly the
same structure.

The observables for which we expect most advantges from the
\MINLO{} method are those that can be constructed from the
momenta of the pseudo-partons after a $\kT$-clustering
procedure carried out until we have $n$ jets, $n$ being
the number of radiated partons beyond the primary process
at the Born level (e.g. $n=1$ for \HJ{} and \ZJ{} and
$n=2$ for  \HJJ{} and \ZJJ{}). In particular, it should work
well for quantities built out of the hardest $n$ jets,
as defined in the inclusive $\kT$ algorithm with a reasonable
(i.e. not too small) choice of the $R$ parameter.
We remark, however, that quantities that are sensitive to the
radiation in the real event (i.e. to the third parton in \HJJ{} and to
the second parton in \HJ{}) the \MINLO{} method has no great advantage
over the standard ones. In fact, no Sudakov suppression is included
for the radiated parton in the real cross section. On the other hand,
the \POWHEG{} method provides specifically these Sudakov form factors,
while maintaining NLO accuracy. Therefore, the \MINLO{} method
combined with \POWHEG{} yields the fully resummed results for
all quantities. We expect that in this
framework the \POWHEG{} results improved with the \MINLO{} method will
ease the task of merging multijet samples, by providing associated jet
cross section that merge more smoothly with those with smaller
multiplicity.

It is possible to conceive observables for which the \MINLO{} method
includes double logarithms (at the NNLO level and beyond) that are
actually not correct \cite{GavinPrivate}. At the end of Section
\ref{sec:HJ} we will consider two such examples.

\subsection{Higgs boson production}\label{sec:H}
\subsubsection{Higgs boson production in association with one jet}\label{sec:HJ}
\begin{figure}[tbh]
\begin{centering}
\includegraphics[width=0.495\textwidth]{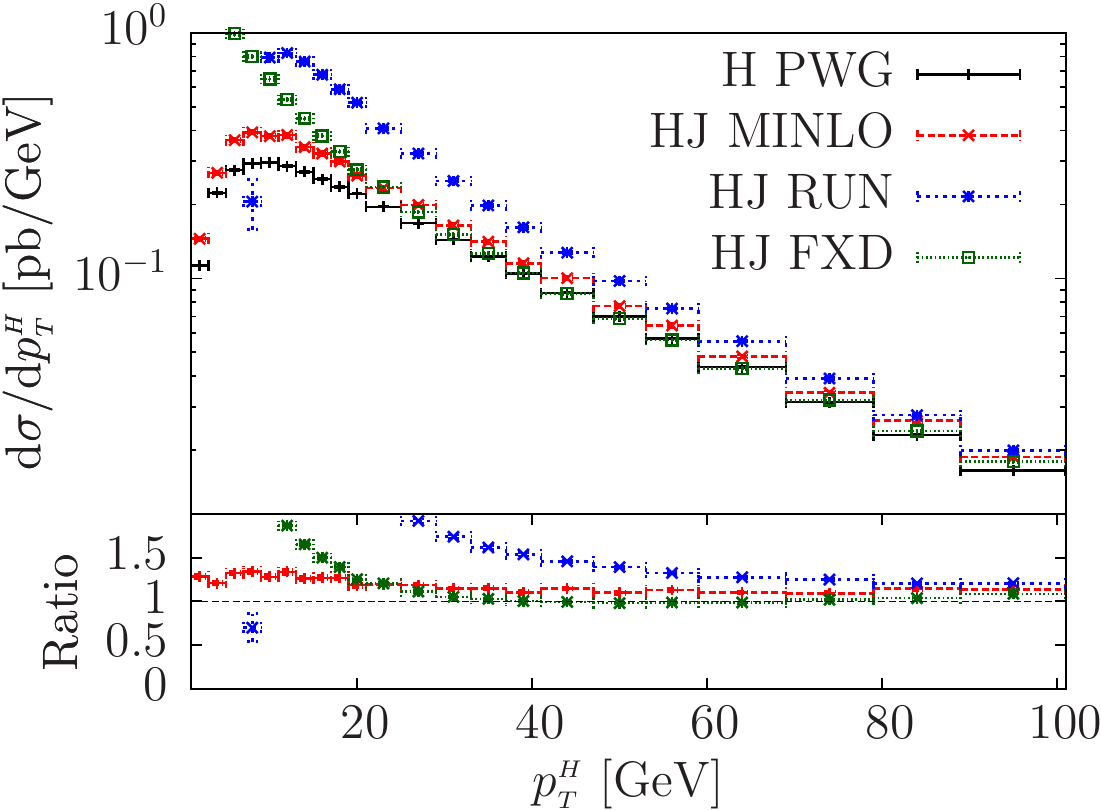}\includegraphics[width=0.477\textwidth]{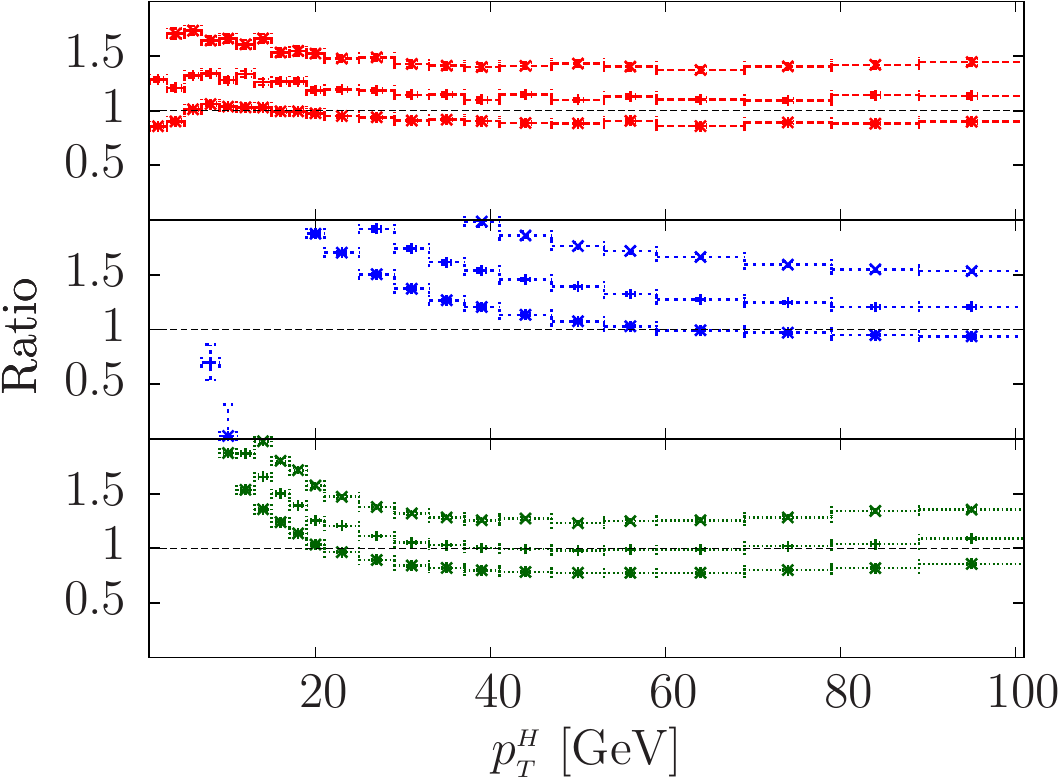} 
\par\end{centering}
\caption{Transverse momentum spectrum of the Higgs boson, computed with the
\POWHEGBOX{} \ggH{} generator (H PWG), the \HJ{}-\MINLO{} result (HJ MINLO), the \HJ{} default
$\muF=\muR=\pTH$ (HJ RUN),
and HJ with $\muF=\muR=\mH$ (HJ FXD). 
The right panel shows the ratio of each of the NLO
HJ results with respect to the NLO \ggH{} \POWHEG{} simulation with
the band either side of the central values indicating the combined
renormalization and factorization scale uncertainty.
Results are shown for LHC collisions
at 7 TeV and a Higgs mass of 120 GeV. No cuts are applied.  
\label{fig:HJ-H-pt}}
\end{figure}

We begin by considering the \MINLO{} improved \HJ{} calculation.  In
fig. \ref{fig:HJ-H-pt}
we show the transverse momentum spectrum of the Higgs boson, computed
with the \POWHEGBOX{} \ggH{} generator, the \HJ{}-\MINLO{} result, and
the \HJ{} result with the two alternative scale choices
$\muF=\muR=\pTH$ (RUN) and $\muF=\muR=\MH$ (FXD).  The
\POWHEGBOX{} result was obtained with the settings advocated in
ref.~\cite{Dittmaier:2012vm}, that is to say with the {\tt hdamp}
parameter set to the Higgs mass divided by 1.2, and all other
parameters at their default value. The events were showered at the
parton level using \PYTHIA{} 6, with the settings described
in the introduction of Section \ref{sec:phenomenology}.

In the right panel in fig.~\ref{fig:HJ-H-pt}, the full scale variation
of the \HJ{} results, normalized to the \POWHEGBOX{} result, are
presented. We have varied the renormalization and factorization scale
independently by a factor of two above and below the central value,
discarding the extreme cases of varying them
in opposite directions.  More precisely, referring
to the notation of Section \ref{sec:method}, we have chosen $\KR$ and
$\KF$ equal to $1/2$, $1$ and $2$, restricted by the condition $1/2\le
\KR/\KF\le 2$.  This leaves seven scales combinations. The upper (lower)
curves are obtained by taking the upper (lower) envelope of these
seven curves.

Notice the striking difference between the \MINLO{} result and the
standard NLO ones. The \MINLO{} result mimics well the \POWHEGBOX{} result
down to very low values of transverse momentum.  We stress again that
this is a consequence of the presence of Sudakov form factors, and
also of the inclusion of the subtraction term of
eq.~(\ref{eq:subtrfac}).  The standard \HJ{} results do instead
diverge at small transverse momentum.  Furthermore, they tend to
abruptly change sign, due to the growing of the large Sudakov double
logarithms arising at the NLO level.  Notice also that they begin to
depart from the \MINLO{} result even at moderate values of the
transverse momenta. By contrast, we observe that the \MINLO{} uncertainty
band is fairly compatible with the \POWHEG{} result down to very low
values of the transverse momentum.

We notice that the fixed scale result is more compatible with the
\MINLO{} result than the running scale one. This may seem surprising,
since, as shown in Section~\ref{sec:further}, the \MINLO{} scale
choice corresponds to the running scale case.  However, the Sudakov
suppression of the \MINLO{} result is missing in the running scale
result. Using a larger scale at small transverse momenta, as is done
in the fixed scale case, compensates to some extent the lack of a
Sudakov form factor, yielding a more stable result.

We also remark that the \MINLO{} result yields an increasing scale
band at low transverse momenta. This is to be contrasted with the
fixed scale case, where the uncertainty band seems to shrink at small
$\pT$, giving the illusion of a smaller theoretical uncertainty. This
observation is easily explained.  The NLO correction includes a
dominant, negative Sudakov term, carrying two more powers of
$\log(\MH/\pT)$ than the Born term. This term causes the NLO
correction to change sign at some small value of $\pT$.  At the point
where the NLO correction becomes zero, its derivative with respect the
renormalization scale almost vanishes. In fact, at the point where the
NLO correction vanishes, the scale dependence is given schematically
by
\begin{equation}
\sigma=B\alphaS^N(\muR)+ N B b_0\log(\muR^2/\mu_0^2) \alphaS^{N+1}(\muR)\,,
\end{equation}
where $\mu_0$ is the scale central value. Its derivative at
$\muR=\mu_0$ is
\begin{equation}
\muR^2\frac{d\sigma}{d \muR^2}=N B\alphaS^{N}(\muR)\left(-b_0\alphaS(\muR)
-b_1\alphaS^2(\muR)\right) + N B b_0 \alphaS^{N+1}(\muR)
=-N B\alphaS^{N+2}(\muR)b_1\;.
\end{equation}
Thus, the scale dependence in this region, being only due to the NLO
evolution term of $\alpha_s$, is small, and the NLO correction is also
small, yielding a full NLO result close to the Born one. Both these
conditions may convey a false impression of reliability.\footnote{ See
  also fig.~3 in ref.~\cite{Bozzi:2008bb}, and the associated
  discussion.} For the \MINLO{} result, on the other hand, this
mechanism does not operate, since the large double logarithmic term is
removed from the NLO correction, and is included in the Sudakov form
factors.

\begin{figure}[tbh]
\begin{centering}
\includegraphics[width=0.487\textwidth]{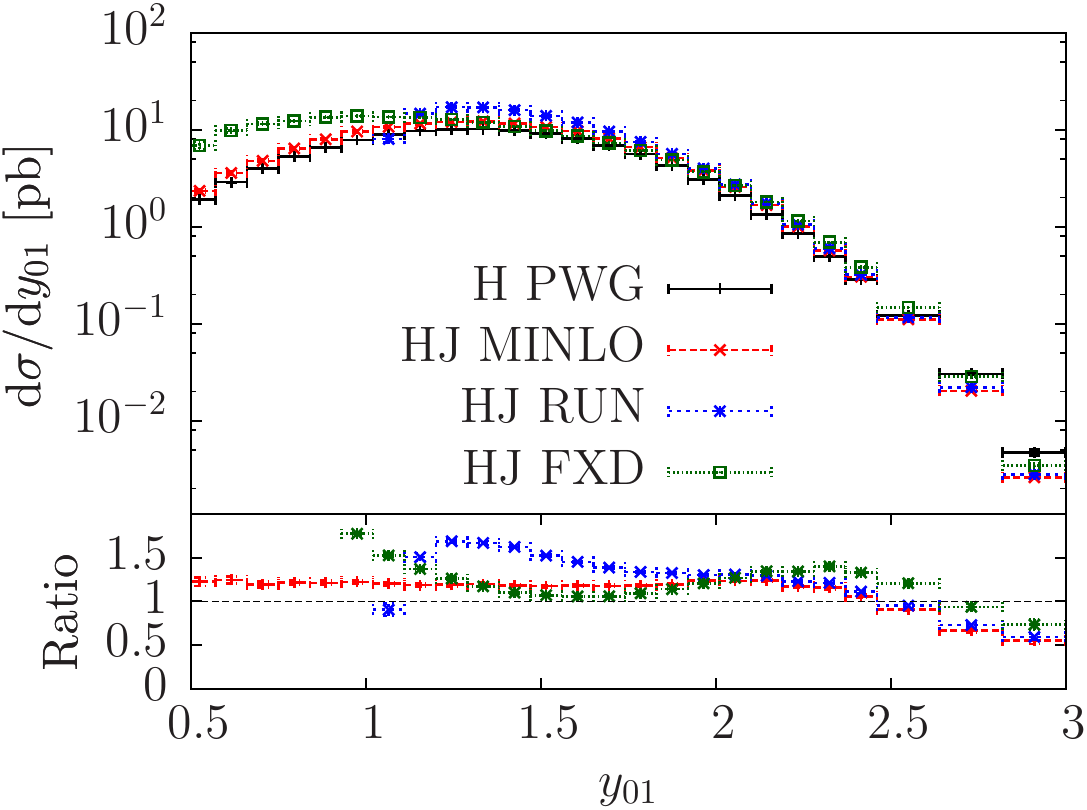} \includegraphics[width=0.495\textwidth]{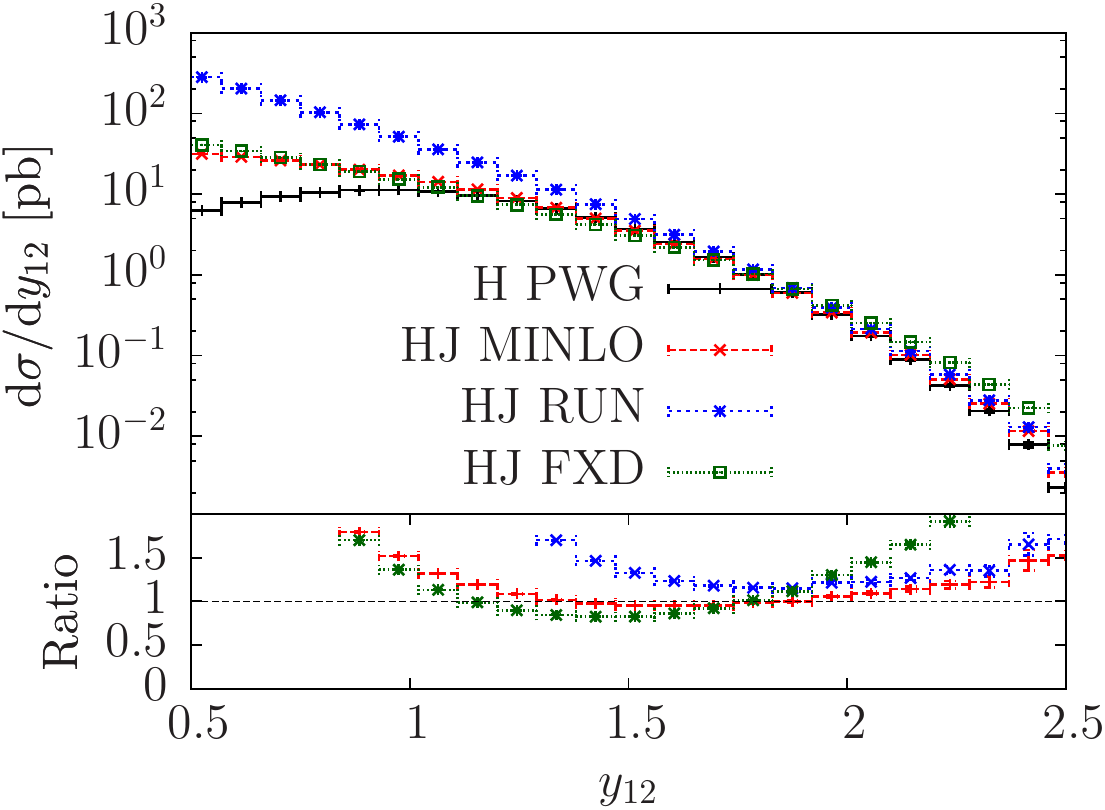} 
\par\end{centering}
\caption{As figure~\ref{fig:HJ-H-pt} for the $0\rightarrow1$ and
  $1\rightarrow2$ differential jet rates in Higgs boson production via
  gluon fusion. 
\label{fig:HJ-yijs} }
\end{figure}

In fig.~\ref{fig:HJ-yijs}
we show the differential jet rates.  These are defined as
\begin{equation}\label{eq:yijdef}
y_{i\,i+1}=\log_{10}\left(\frac{q_{i\,i+1}}{1\,{\rm GeV}}\right),
\end{equation}
where $q_{i\,i+1}$ is the $\kT$ merging scale for going from a
$(i+1)$-jet configuration to a $i$-jet configuration in the $\kT$
clustering procedure.  For the $0\rightarrow 1$ differential jet rate,
as for the Higgs transverse momentum distribution, all three methods
we are considering should be predictive.  We see again that the
\MINLO{} prediction is well behaved even below $q_{01}\approx
10\;$GeV, while the standard methods fail in this region.  The
$1\rightarrow 2$ differential jet rate is a distribution for which we
expect little or no improvement from the \MINLO{} method. In fact, it
is determined by the distribution of the radiated parton in real
events, that forms the second jet. No Sudakov improvement for this
emission is provided by the \MINLO{} method.  In the \POWHEG{} Higgs
implementation, this quantity is determined by the shower stage of the
generation, where partons beyond the first one are generated.  In a
\POWHEG{} simulation of the \HJ{} process, either with the
\MINLO{} improvement or with a standard choice of scales, all curves
would be in better agreement with the \POWHEG{} \ggH{} result, since
in this case a Sudakov form factor for the radiated parton is properly
included.

\begin{figure}[tbh]
\begin{centering}
\includegraphics[width=0.495\textwidth]{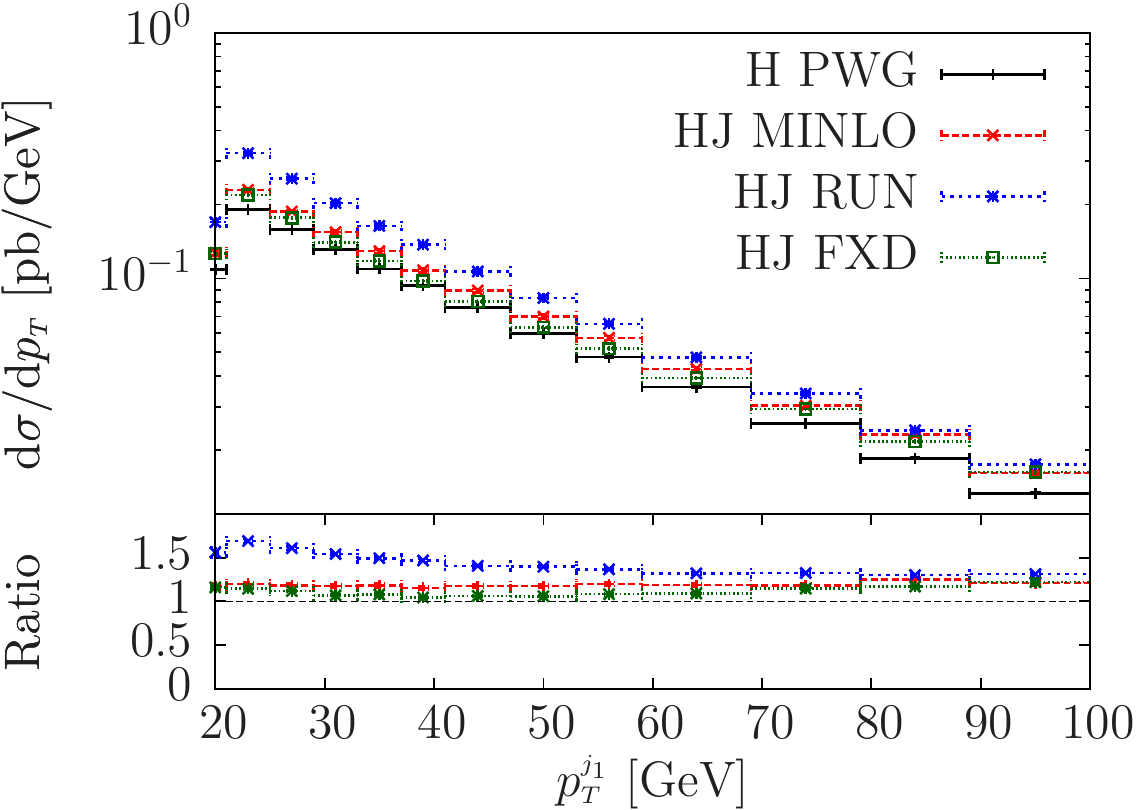}
\includegraphics[width=0.495\textwidth]{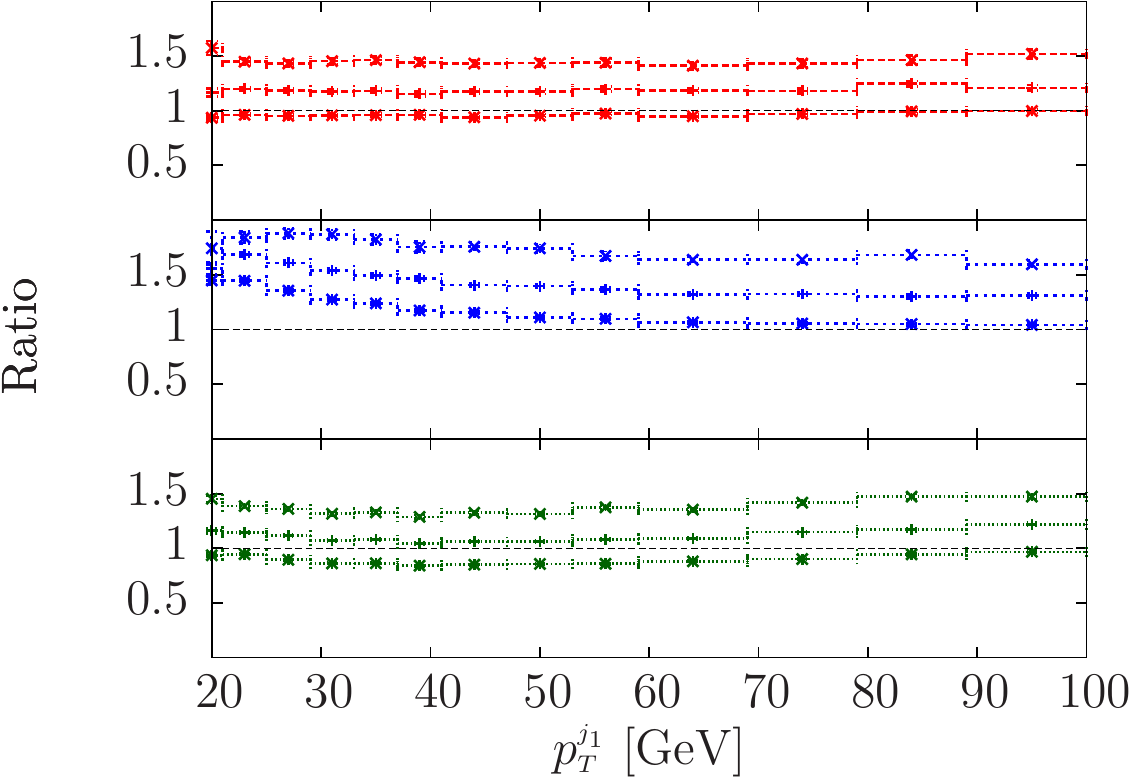} 
\par\end{centering}
\caption{As figure~\ref{fig:HJ-H-pt} for the transverse momentum
  spectrum of the leading jet produced in Higgs boson production via
  gluon fusion.
\label{fig:HJ-j1-pt}}
\end{figure}

In fig.~\ref{fig:HJ-j1-pt}
the transverse momentum of the hardest jet is plotted (this
distribution is equivalent to $y_{01}$ but shown in a different form
and range).
 This distribution should be similar to the transverse momentum of the
 Higgs, except for the fact that it is displayed for $p_T > 20$~GeV.
 As the small $\pT$ region is approached, we see indications of an
 initial unphysical behaviour in the standard methods, especially
 evident for the running scale case.

\begin{figure}[tbh]
\begin{centering}
\includegraphics[width=0.495\textwidth]{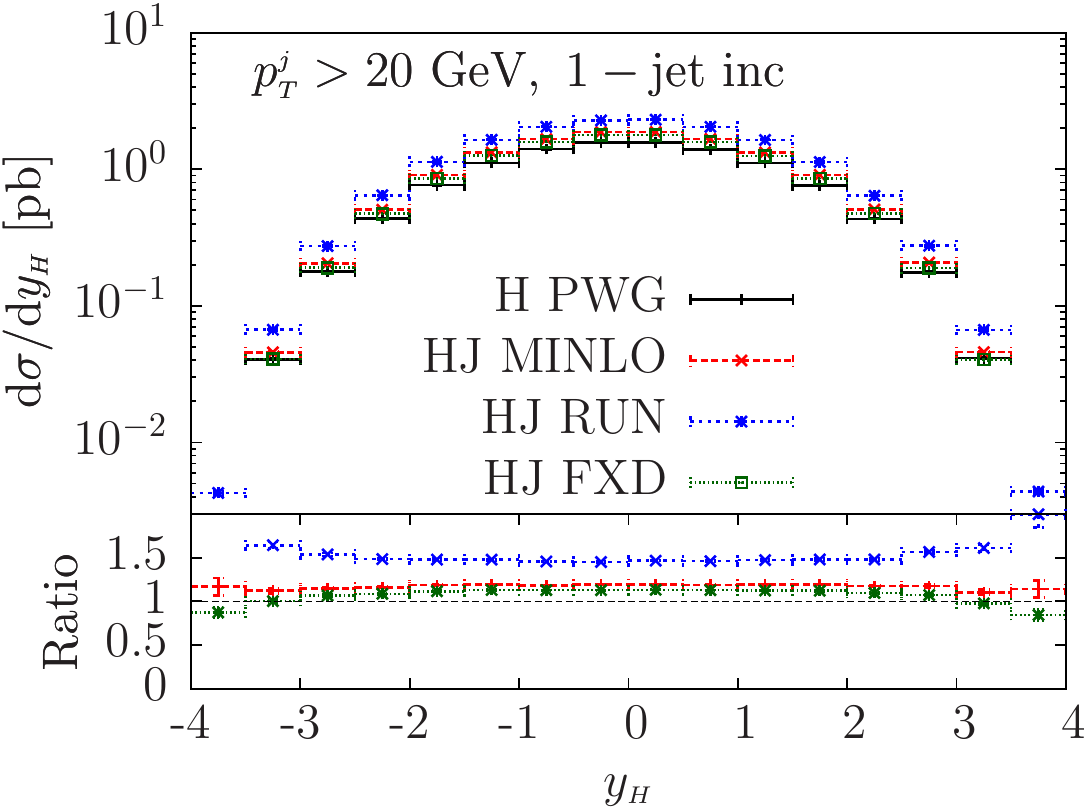}
\includegraphics[width=0.477\textwidth]{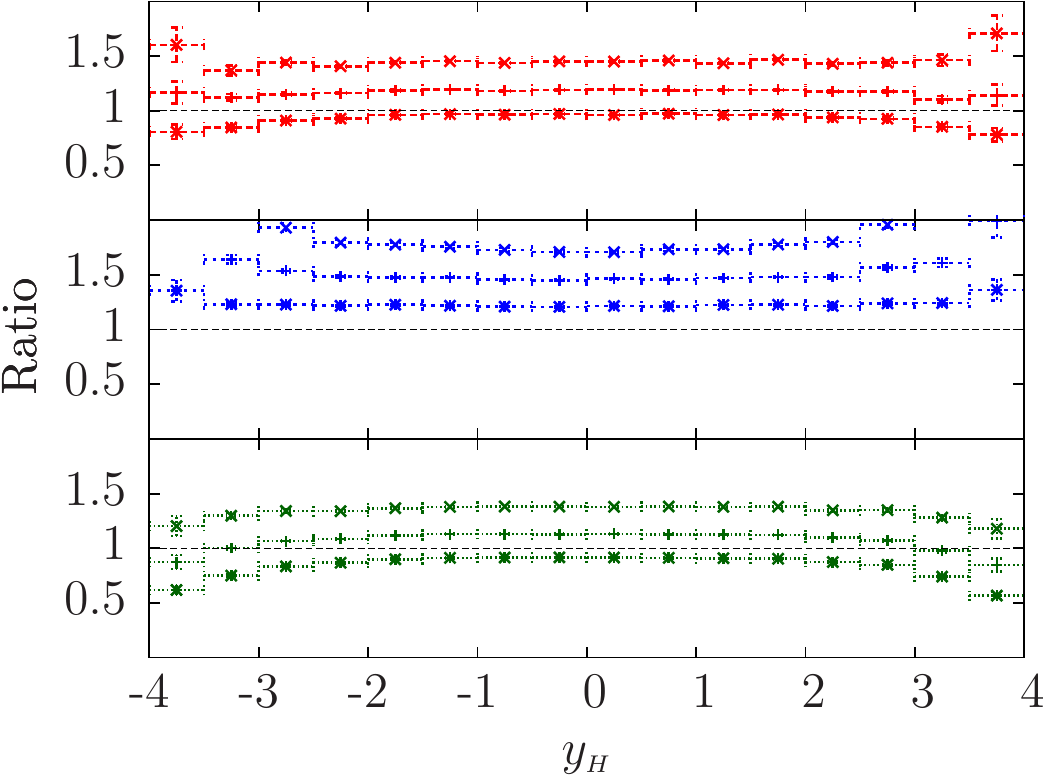} 
\par\end{centering}
\caption{
As figure~\ref{fig:HJ-H-pt} for the rapidity of the Higgs boson in
events containing at least one jet with $p_T^j > 20$ GeV. 
\label{fig:HJ-y-1j-inc} }
\end{figure}

In fig.~\ref{fig:HJ-y-1j-inc}
we show the rapidity distribution of the Higgs, in events with at
least one jet above 20~GeV.  The normalization of this distribution
inherits the results obtained for the jet transverse momentum
distribution, since it is mainly affected by the 20 GeV cut. An
interesting trend is observed, however, in the large rapidity region,
where phase space restrictions become operative and Sudakov effects
may become manifest. The \MINLO{} result is more compatible with the
\POWHEG{} \ggH{} one, while the default running scale result markedly
departs from it.

We consider now two distributions such that the double logarithmic
structure introduced with the \MINLO{} procedure is not correct beyond
the NLO level~\cite{GavinPrivate}. Consider for example the transverse
momentum distribution of the hardest jet included in a given rapidity
range around the Higgs boson and of the hardest jet in a fixed
(central) rapidity region.  Observe that in both cases the jet in the
considered rapidity range is not necessarily the hardest one in the
whole process, since the hardest jet could be outside that range.
Thus, these distributions do not satisfy our requirement mentioned
earlier, i.e. they cannot be constructed neither out of the inclusive
$\kT$-clustered configuration at the one-jet level, nor out of the
inclusive hardest jet distribution. Since we are limiting the
collinearity of the emitted jet, the \MINLO{} procedure introduces an
excessive double logarithmic Sudakov suppression for these
observables.
\begin{figure}[tbh]
\begin{centering}
\includegraphics[width=0.495\textwidth]{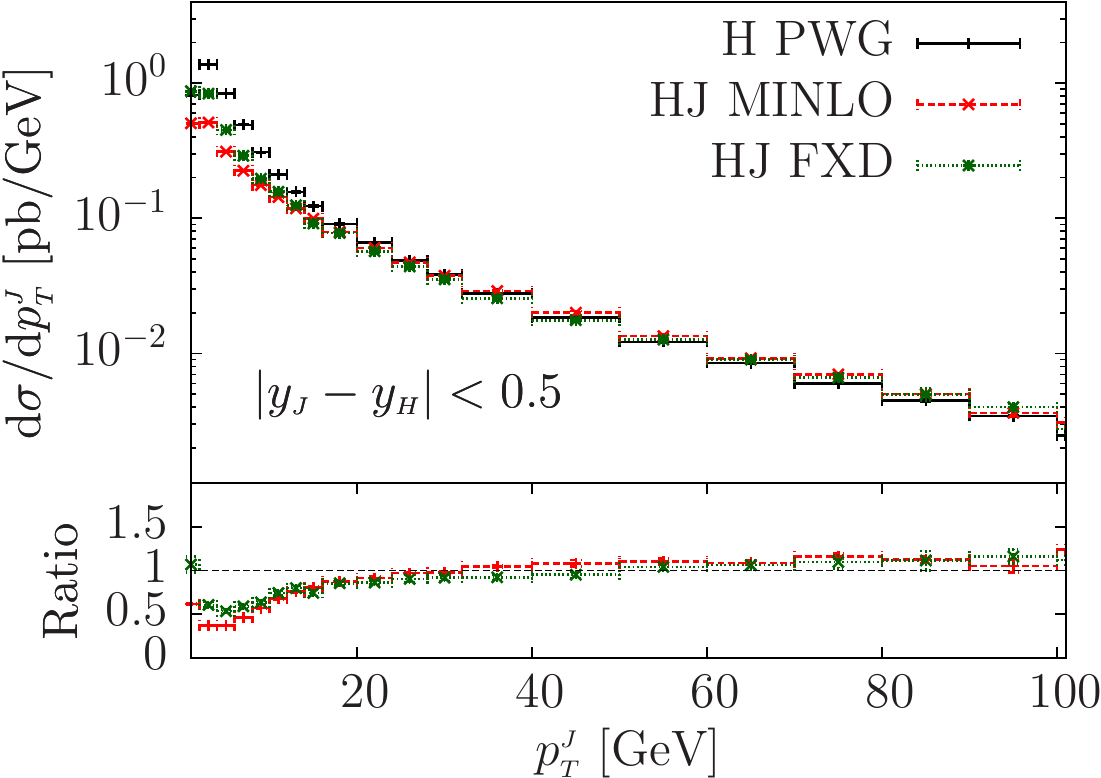}
\includegraphics[width=0.477\textwidth]{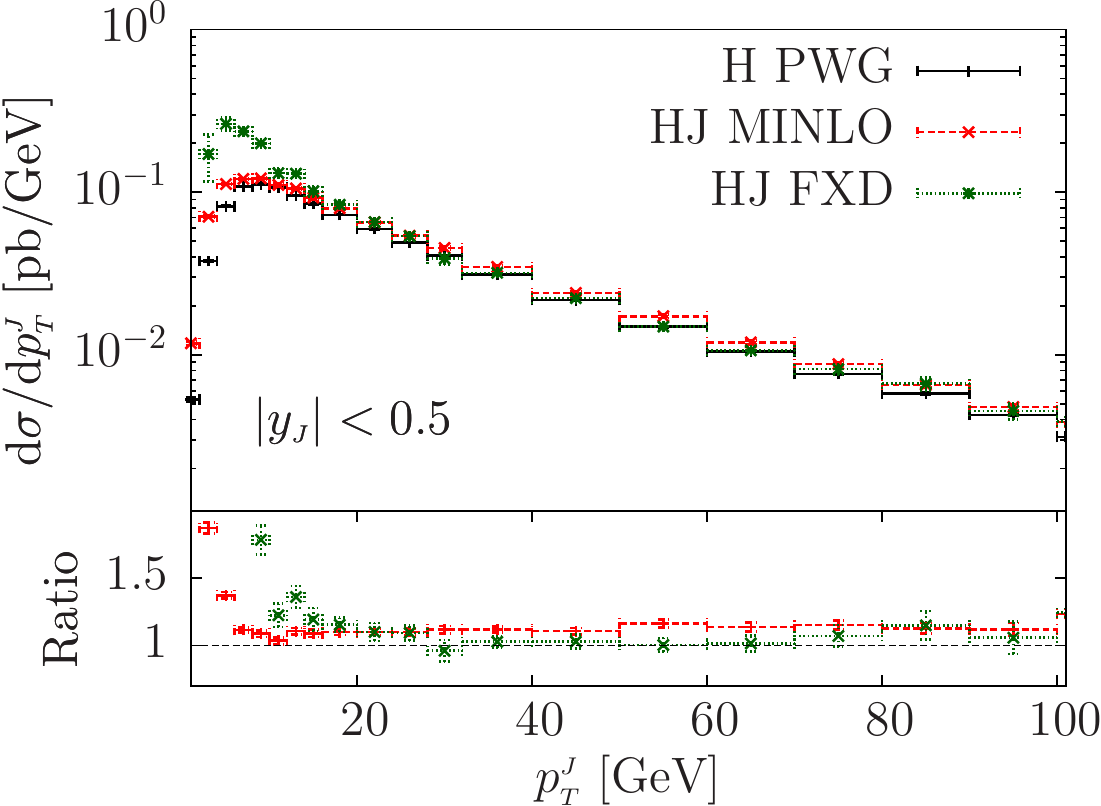} 
\par\end{centering}
\caption{
The transverse momentum distribution of the hardest with $|y_J-y_H|<0.5$
(left plot) and with $|y_J|<0.5$ (right plot).
\label{fig:gavin} }
\end{figure}
In fig.~\ref{fig:gavin} we show predictions for the transverse
momentum distribution of the hardest jet with $|y_J-y_H|<0.5$ (left
plot) and $|y_J|<0.5$ (right plot) from, the showered \ggH{}
generator, the \MINLO{} procedure and a standard NLO calculation with
a fixed scale. We observe that indeed for the left plot the \MINLO{}
prediction seems to fare worse than the standard fixed order NLO
computation.  On the other hand, relevant differences between the two
are only visible for very small $\pT$, already in the region where
both distributions depart from the \ggH{} prediction. Conversely, for
the right plot the \MINLO{} prediction tracks the \ggH{} one more
closely down to smaller values of $\pT$. Even for the left plot, the
\MINLO{} result is closer to the \ggH{} one as soon as one increases
the rapidity interval. We thus see that even for observables conceived
to expose the limitations of the \MINLO{} method, it still performs
comparably to standard NLO calculations.  Nevertheless, we believe
that more extensive experience of the \MINLO{} method is needed in
order to fully assess its performance.

\subsubsection{NLO Higgs boson production in association with two jets\label{sub:NLO-HJJ}}
\begin{figure}[tbh]
\begin{centering}
\includegraphics[width=0.495\textwidth]{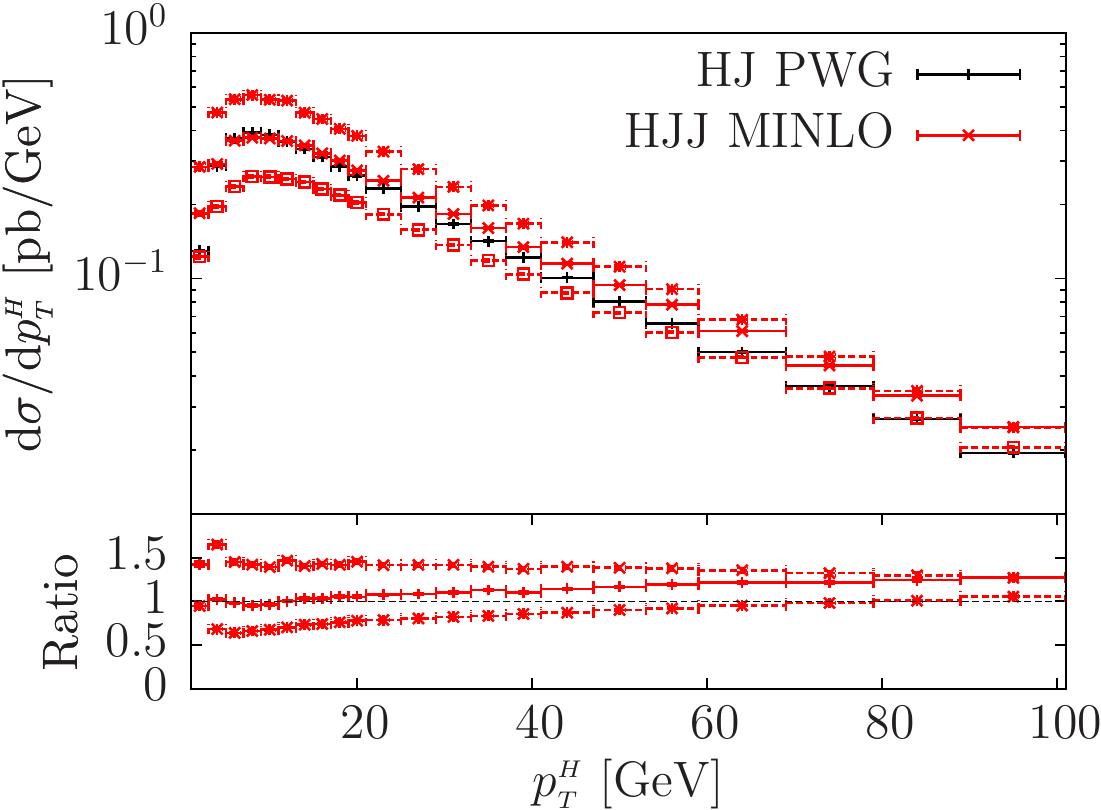} \includegraphics[width=0.488\textwidth]{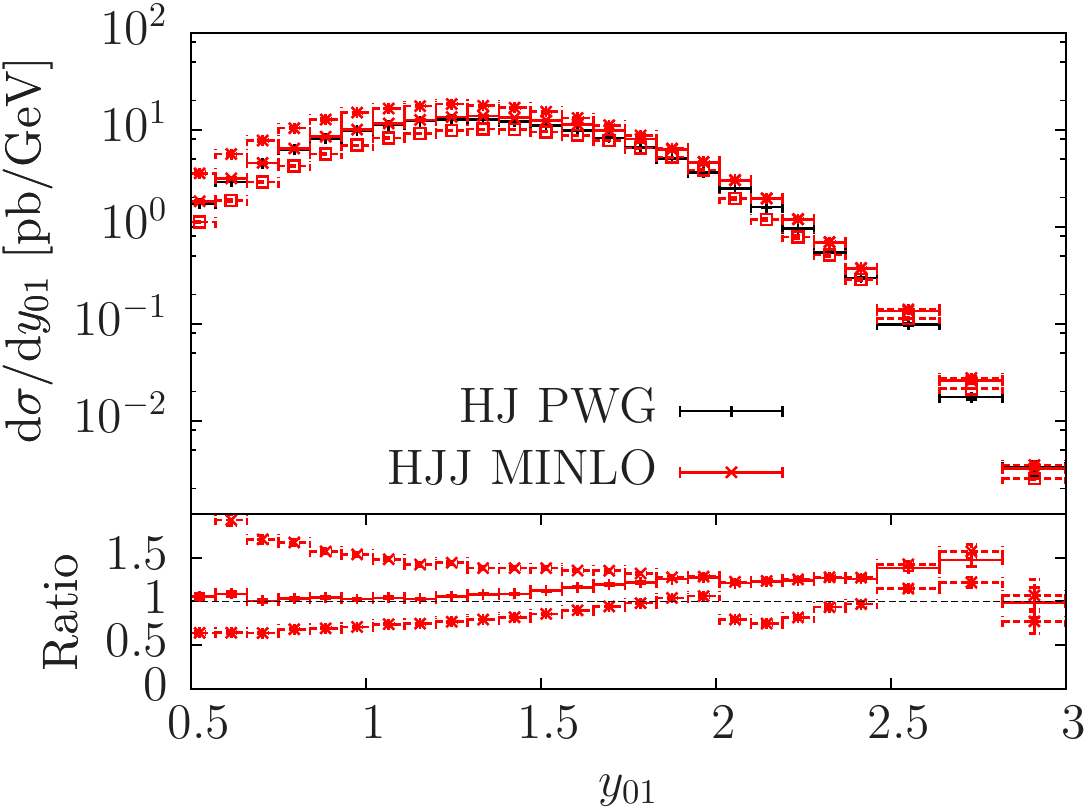} 
\par\end{centering}
\caption{The transverse momentum of the Higgs boson (left) and the
  differential jet rate $y_{01}$ (right), representing the logarithm
  of the resolution scale in the $\kT$ jet algorithm
  \cite{Catani:1993hr} for which 1-jet events become resolved as 0-jet
  ones.  Results shown are computed with the \POWHEGBOX{} \HJ{}
  generator, augmented by the \MINLO{} procedure, and with
  \HJJ-\MINLO{} method.
  Distributions are shown for LHC collisions at 7 TeV and a Higgs mass of
  120 GeV. No cuts are applied.
  \label{fig:HJJ-ptzoom-inc-and-y01} }
\end{figure}

In this section we compare the \MINLO{} \HJJ{} distributions with the
standard ones, obtained with two choices of the scale,
$\muF=\muR=\mH$, and $\muF=\muR=\HThat$, with $\HThat$ defined as
\begin{equation}
\HThat=\sqrt{{\mH}^2+{\pTH}^2}+\sum_i \pT^{(i)},
\end{equation}
the sum running on all final state partons. These two scale choices
will be labelled FXD and RUN in the figures.  We begin by
comparing in fig.~\ref{fig:HJJ-ptzoom-inc-and-y01}
the transverse momentum of the Higgs obtained with the \POWHEGBOX{}
\HJ{} generator (interfaced to the \PYTHIA{} shower) and the \MINLO{}
\HJJ{} generator.  The \POWHEGBOX{} \HJ{} generator was modified with
the inclusion of the \MINLO{} method for the computation of the
underlying Born kinematics. No standard NLO Higgs plus two jets
prediction is possible for this distribution, since it does not
require the presence of at least two jets. Thus, as previously
discussed, we expect the \MINLO{} result to give a LO representation
of the physical cross section. We can see that, in spite of this the
\MINLO{} result is still remarkably close to the \POWHEGBOX{} cross
section. The agreement is particularly impressive at very low
transverse momentum, where it seems that the \MINLO{} \HJJ{} result
gives a description of the total Higgs cross section that is very
close to the one given by the \HJ{} \POWHEGBOX{} generator. The
latter, when improved with the \MINLO{} prescription, yields a cross
section that is accurate at least at LO, according to the discussion
given at the beginning of Section \ref{sec:phenomenology}.  In the
right panel of fig.~\ref{fig:HJJ-ptzoom-inc-and-y01} we show the
differential jet rate for the zero jet to one jet transition. Here 
again we see the \MINLO{} prediction closely tracks the result of
the \HJ{} \POWHEG{} generator.

\begin{figure}[tbh]
\begin{centering}
\includegraphics[width=0.495\textwidth]{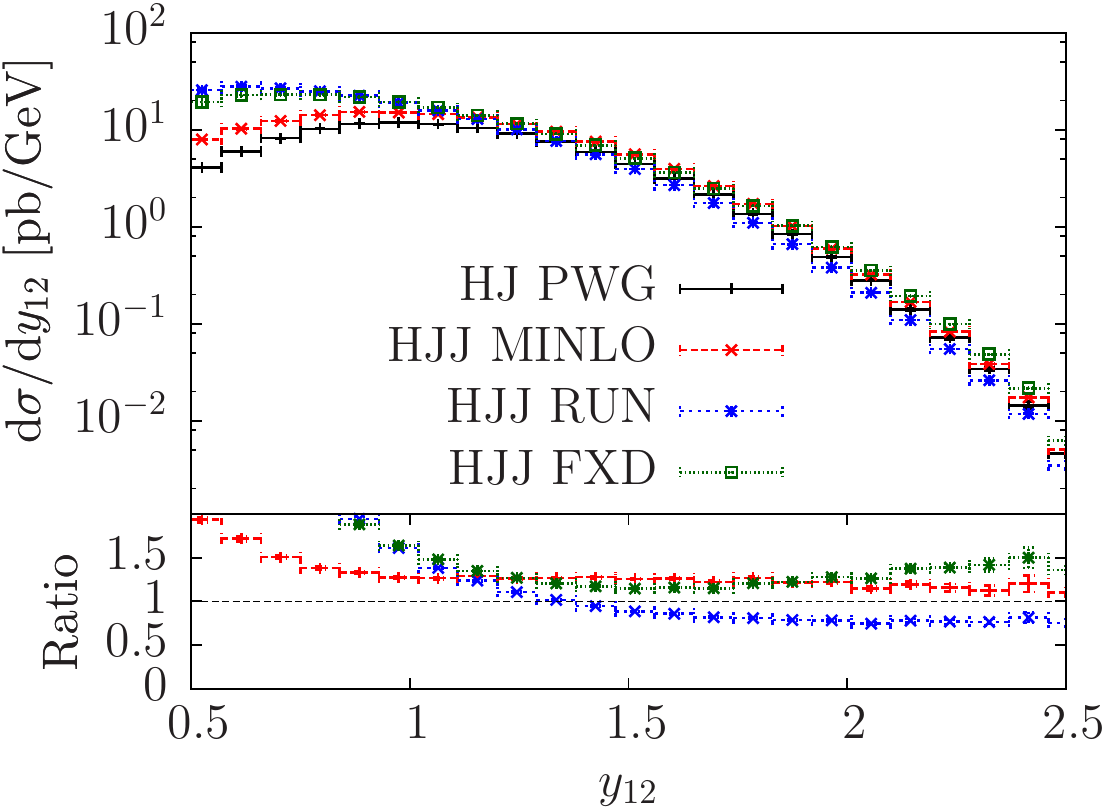} \includegraphics[width=0.477\textwidth]{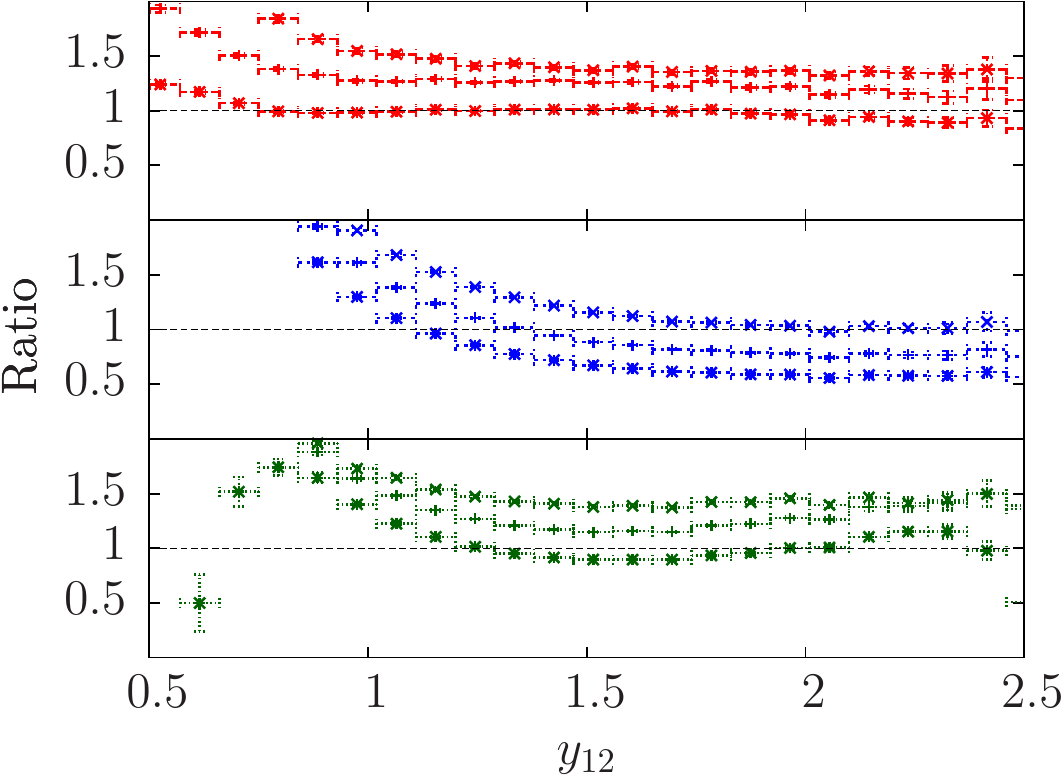} 
\par\end{centering}
\caption{
The differential jet rate $y_{12}$, defined as the value of the $k_{T}$
jet measure \cite{Catani:1993hr} for which events with two resolved
jets are clustered into 1-jet events.
Results are computed with the
\POWHEGBOX{} \HJ{} generator (HJ PWG), the \HJJ{}-\MINLO{} result (HJJ
MINLO), the \HJJ{} with 
$\muF=\muR=\hat H_T$ (HJJ RUN),
and HJJ with $\muF=\muR=\mH$ (HJJ FXD). 
To the right we show the ratio of each of the NLO
HJJ results with respect to the NLO \HJ{} \POWHEG{} simulation, with
the band either side of the central values indicating the combined
renormalization and factorization scale uncertainty.
\label{fig:HJJ-y12} }
\end{figure}

In fig.~\ref{fig:HJJ-y12}
the differential jet rate $y_{12}$ is shown. For this distribution the
\MINLO{} result and the standard NLO calculations are all predictive,
showing reasonable agreement among each other for moderately large
merging scales. At small scales, the \MINLO{} result is in better
agreement with the \POWHEGBOX{} \HJ{} code and shows a better
scale stability. The standard \HJJ{} NLO results, by constrast,
display unphysical behaviour under scale variation, especially as far
as the $\HThat$ scale choice is concerned.

\begin{figure}[tbh]
\begin{centering}
\includegraphics[width=0.495\textwidth]{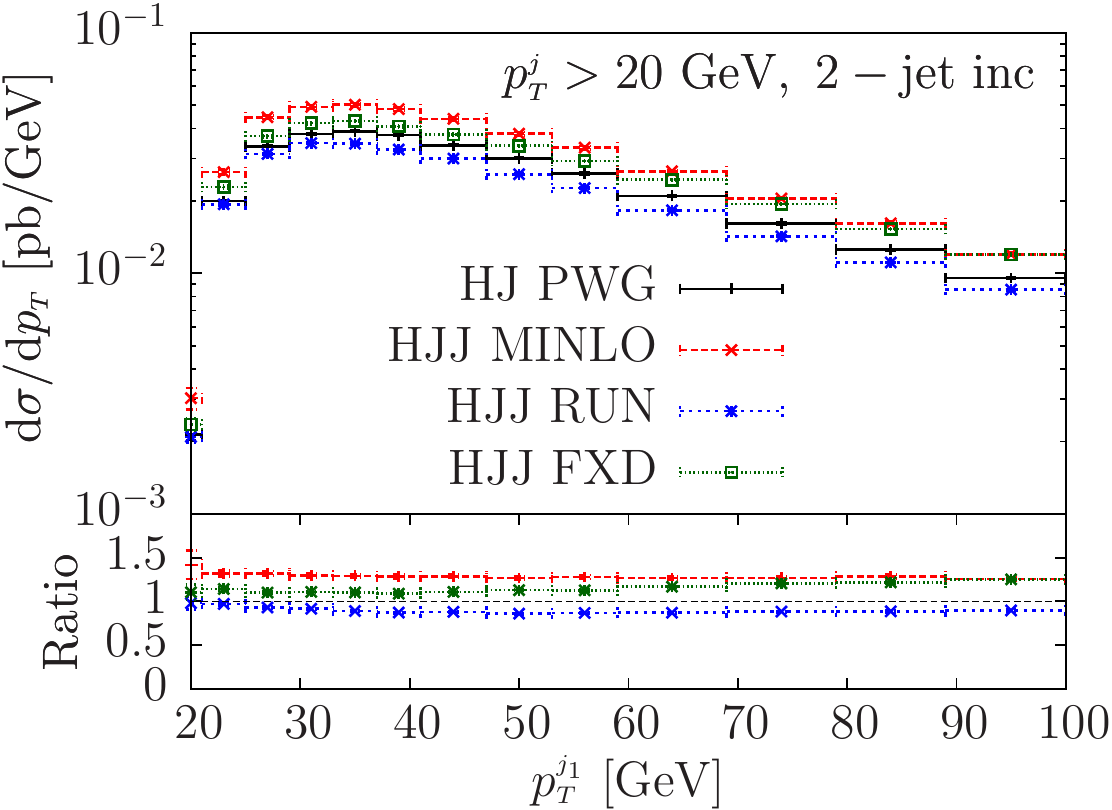}
\includegraphics[width=0.477\textwidth]{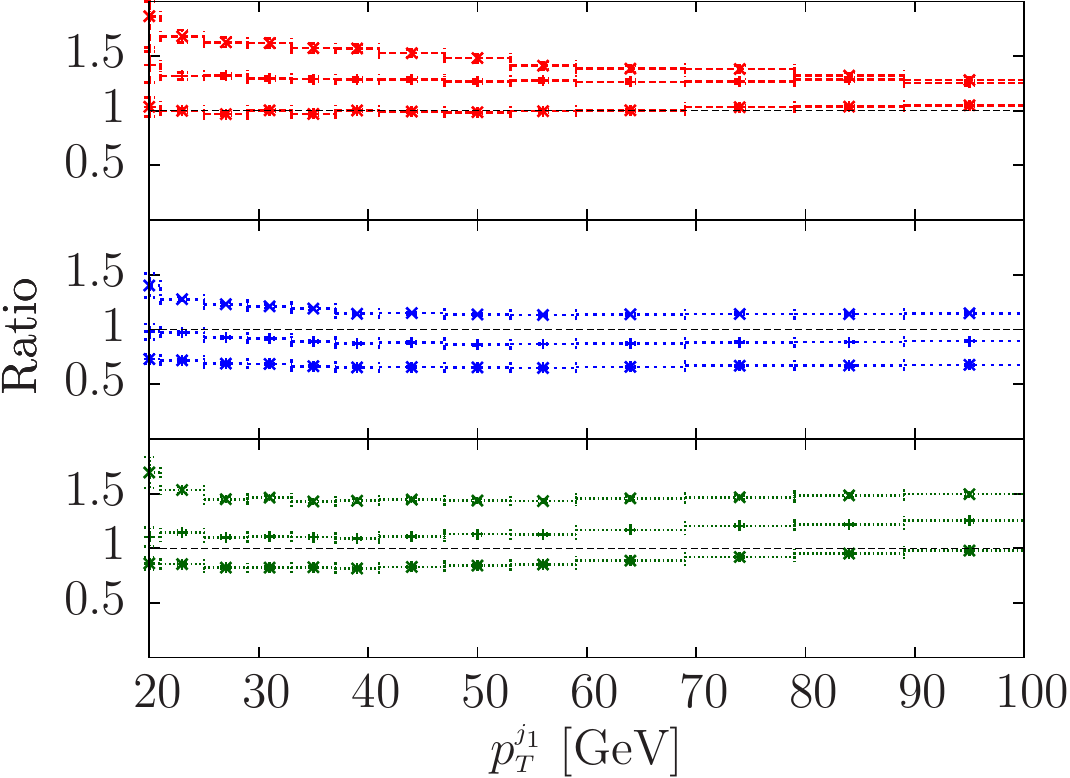} 
\par\end{centering}
\caption{ As in fig.~\ref{fig:HJJ-y12}, for the transverse momentum of
  the leading jet in events with a Higgs boson and at least two jets
  with $\pT>20$ GeV.\label{fig:HJJ-Njge2-j1-pt-020}
}
\end{figure}

In fig. \ref{fig:HJJ-Njge2-j1-pt-020}
we show the transverse momentum of the leading jet in events with at
least two jets. All NLO calculations, \MINLO{}-improved and those with
conventional scale setting, are again predictive for this distribution.
Observe that in the case of the running scale prediction ($\muR=\muF=\HThat$)
the central value is outside the \MINLO{} error band.
Using a central value of $\HThat/2$ would instead lead to 
much better agreement between the \MINLO{} and RUN results. Remarkably, it
has become common in multijet NLO calculation to prefer $\HThat/2$ as
central scale, because it seems to lead to an improved scale
stability. The \MINLO{} result seems also to favour this choice.  We
also notice that the uncertainty band for the \MINLO{} result shrinks
at high $\pT$, while those of the NLO results using a more
conventional scale choice do not. It is tempting to interpret this
result as being due to the fact that the \MINLO{} method yields
smaller radiative corrections in the high $\pT$ region, on account
of its resummation of logarithms of the ratio of the widely different
scales present in this observable --- the jet $\pT$ cut and
$\pT$ of the first jet.  On the other hand, we must remember that the terms
that are exponentiated in the Sudakov form factor are not subject to
scale variation in our present procedure. Thus, we believe that much
more practice with \MINLO{} calculations is needed in order to
substantiate this interpretation.

\begin{figure}[tbh]
\begin{centering}
\includegraphics[width=0.495\textwidth]{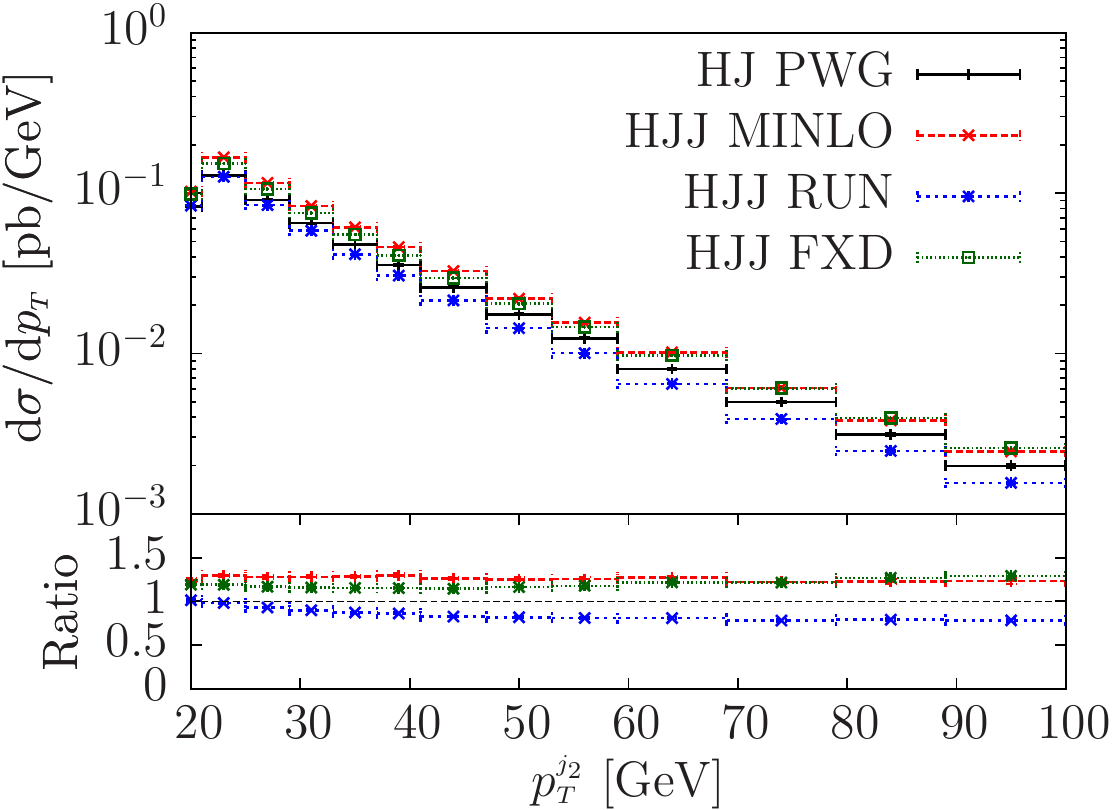}
\includegraphics[width=0.477\textwidth]{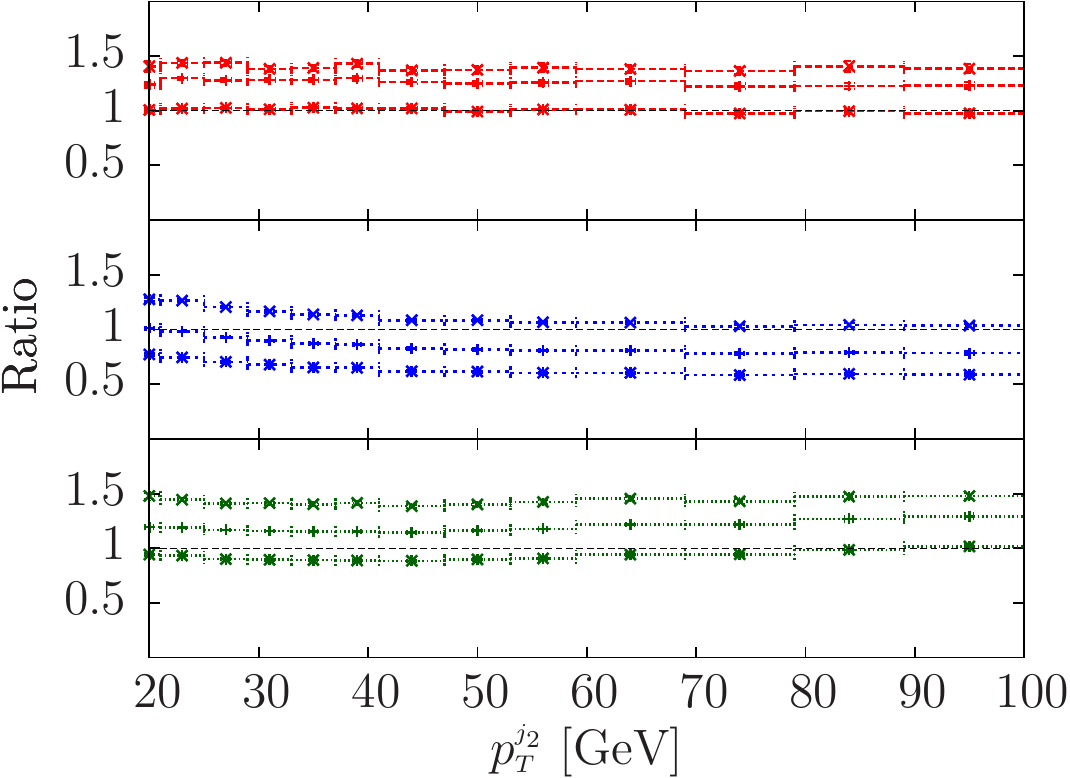} 
\par\end{centering}
\caption{ As in fig.~\ref{fig:HJJ-y12}, for the transverse momentum of
  the next-to-leading jet in events with a Higgs boson and at least
  two jets with $\pT>20$ GeV. \label{fig:HJJ-j2-pt-020}
}
\end{figure}

In fig.~\ref{fig:HJJ-j2-pt-020}
we show the transverse momentum distribution of the second jet. The
$\HThat$ scale choice gives results below the \MINLO{} ones, to an
even larger extent than in the case of the leading jet. Again,
choosing $\HThat/2$ as central scale considerably improves the
agreement, although not quite in a satisfactory way. Predictions with
$\HThat/2$ remain on the lower limit of the \MINLO{} band for moderate
transverse momenta.

\subsection{$Z$ boson production}\label{sec:Z}

\subsubsection{$Z$ boson production in association with one jet}\label{sec:ZJ}
\begin{figure}[tbh]
\begin{centering}
\includegraphics[width=0.495\textwidth]{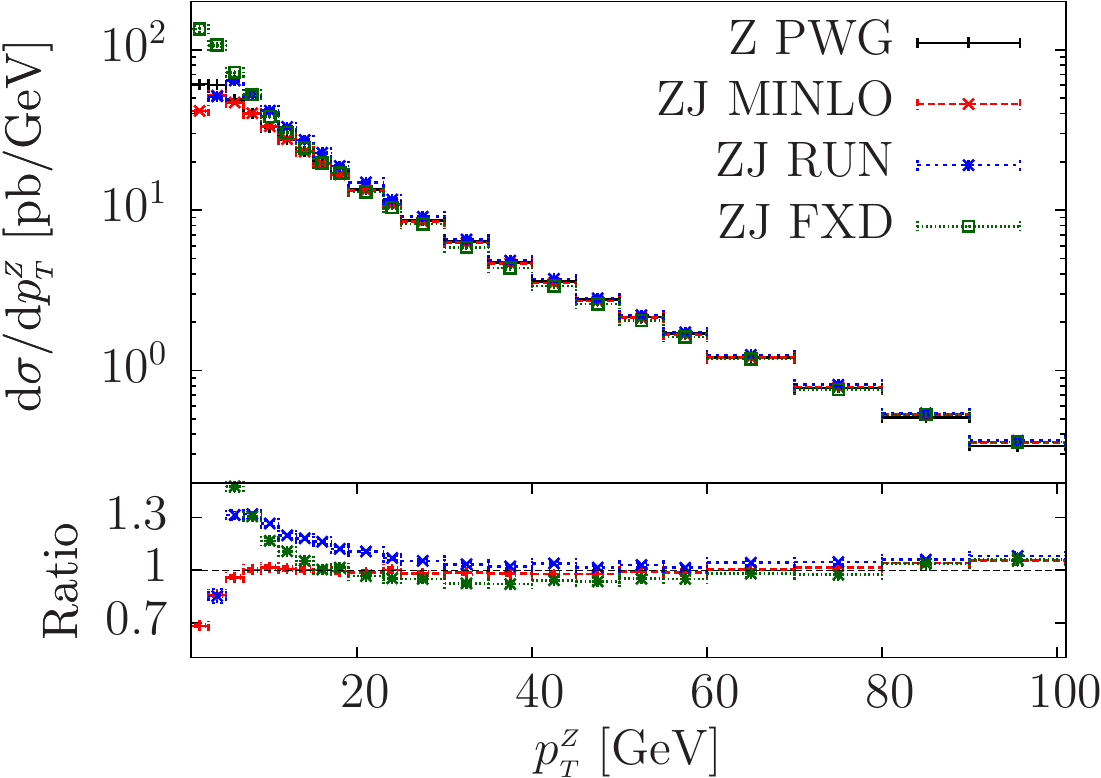}
\includegraphics[width=0.477\textwidth]{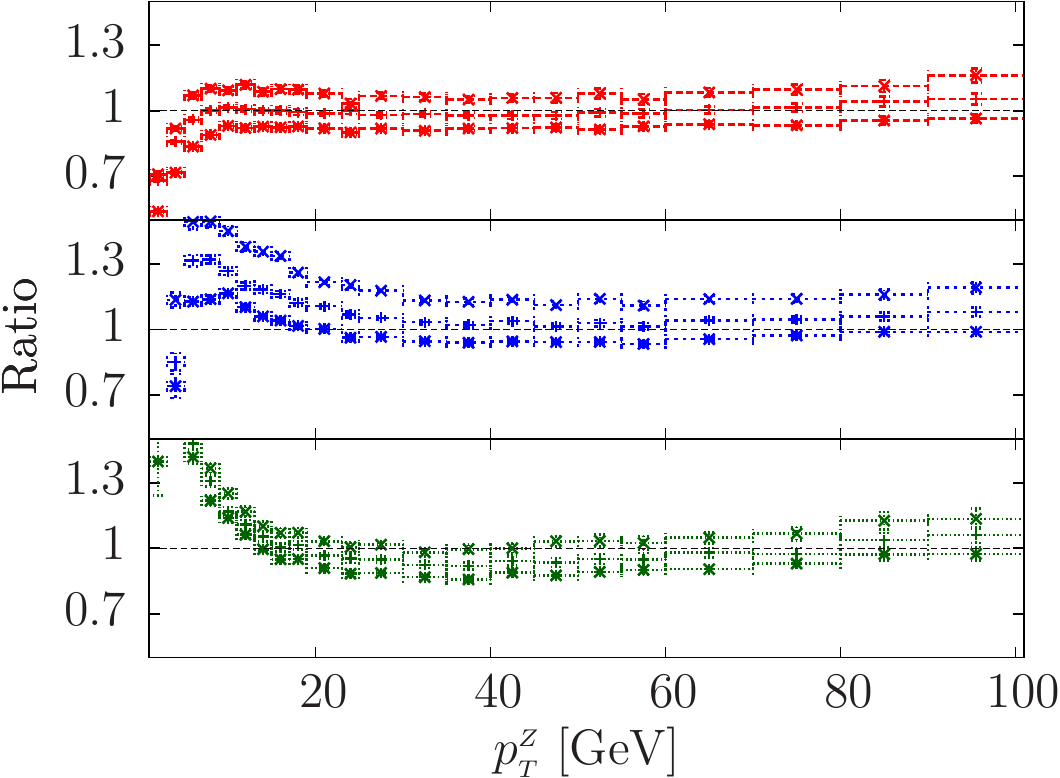} 
\par\end{centering}
\caption{Transverse momentum spectrum of the Z boson computed with
  the \POWHEGBOX{} Z generator, the ZJ-MINLO result, the ZJ
  default $\muF=\muR=\pTZ$ (ZJ RUN) and ZJ with $\muF=\muR=\mZ$ (ZJ
  FXD).  To the right we show the ratio of each of the NLO \ZJ{}
  results with respect to the NLO \Z{} \POWHEG{} simulation, the
  band either side of the central values indicating the combined
  renormalization and factorization scale uncertainty.
\label{fig:ZJ-Z-pt}}
\end{figure}

We begin by showing the transverse momentum of the
$Z$ boson in fig.~\ref{fig:ZJ-Z-pt},
wherein we compare the predictions of the \POWHEG{} $Z$ program interfaced
to \PYTHIA{}, \MINLO{} \ZJ{} and conventional NLO $Z+\mathrm{jet}$
computations using two different scale choices: the mass of the Z boson
and, separately, its transverse momentum. Scale uncertainty bands are
presented in the accompanying plot to the right of the main distribution.
As one
can see, the \MINLO{} result is closer to the \POWHEG{} one for small
transverse momenta.  Observe, however, that now the agreement with
\POWHEG{} is not as good as in the Higgs case. It should be kept in
mind, however, that the $Z$ Sudakov peak is located at much smaller
values of the transverse momentum with respect to the Higgs case, in a
region that is strongly influenced by cut-offs introduced to avoid
the Landau pole in the perturbative calculation, and by shower
cut-offs.

\begin{figure}[tbh]
\begin{centering}
\includegraphics[width=0.495\textwidth]{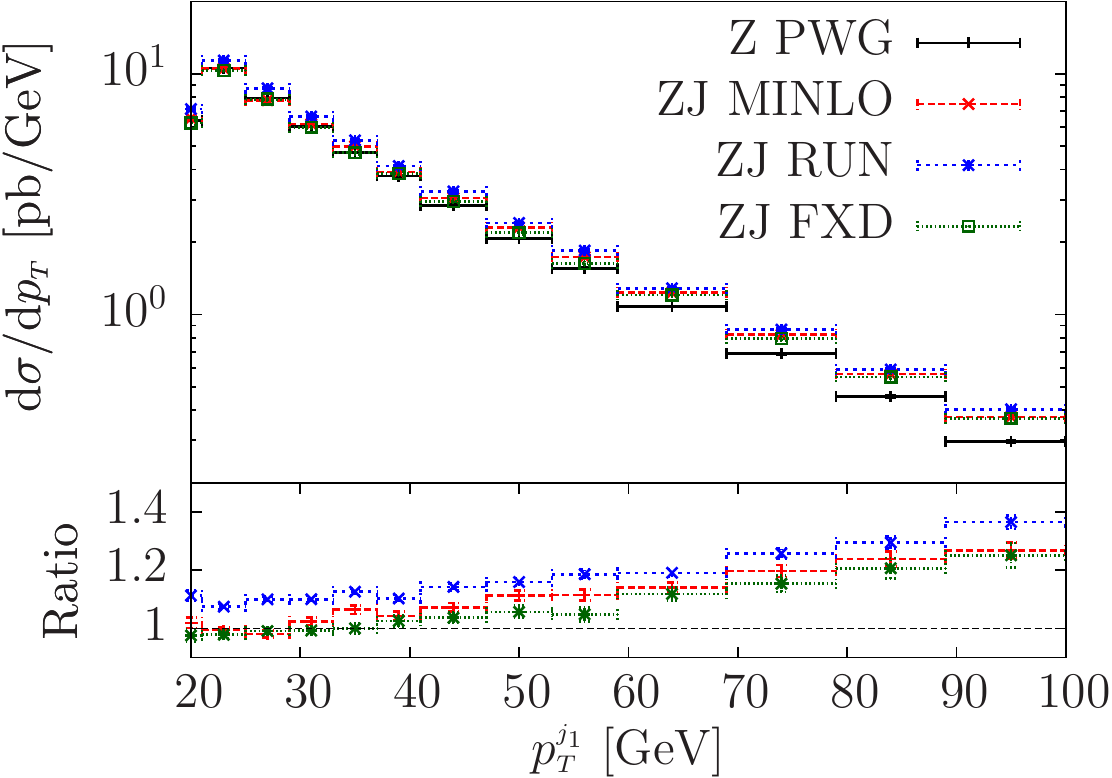}
\includegraphics[width=0.477\textwidth]{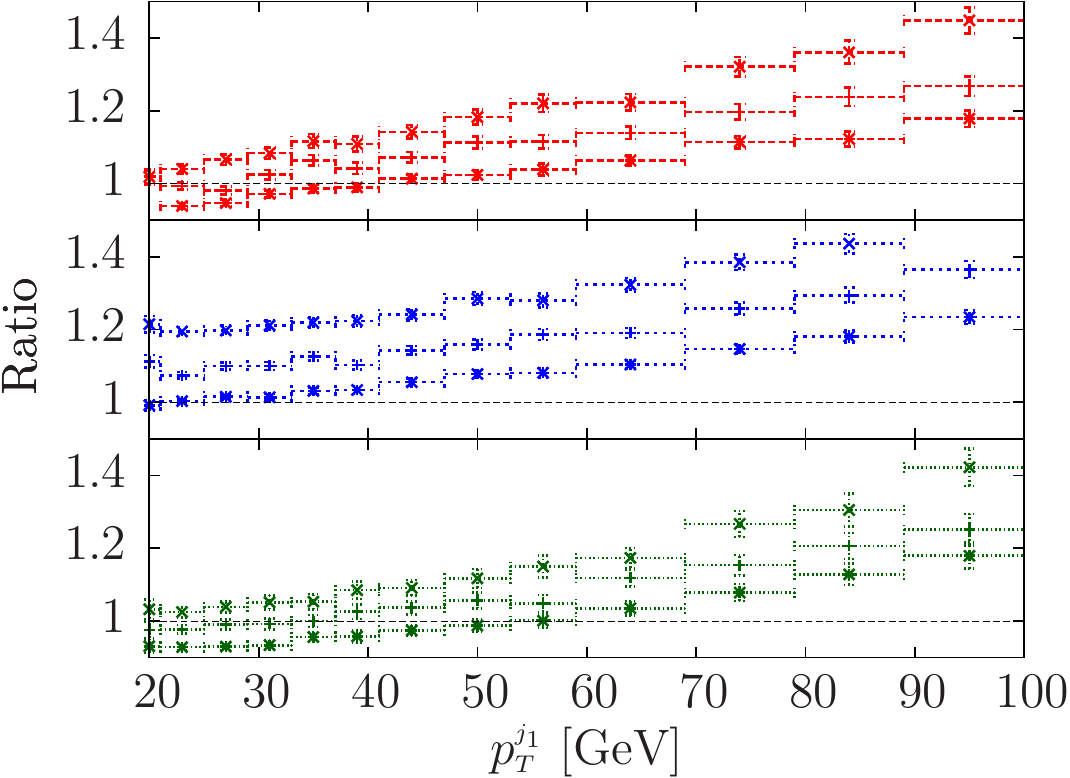} 
\par\end{centering}
\caption{ As figure~\ref{fig:ZJ-Z-pt} for the transverse momentum
  spectrum of the leading jet produced in Z boson
  production. \label{fig:ZJ-j1-pt}}
\end{figure}

In fig.~\ref{fig:ZJ-j1-pt} we display the hardest jet transverse
momentum. In this case, due to the 20 GeV cut on the jet transverse
momentum, we see that all prescriptions perform equally well.

\begin{figure}[tbh]
\begin{centering}
\includegraphics[width=0.495\textwidth]{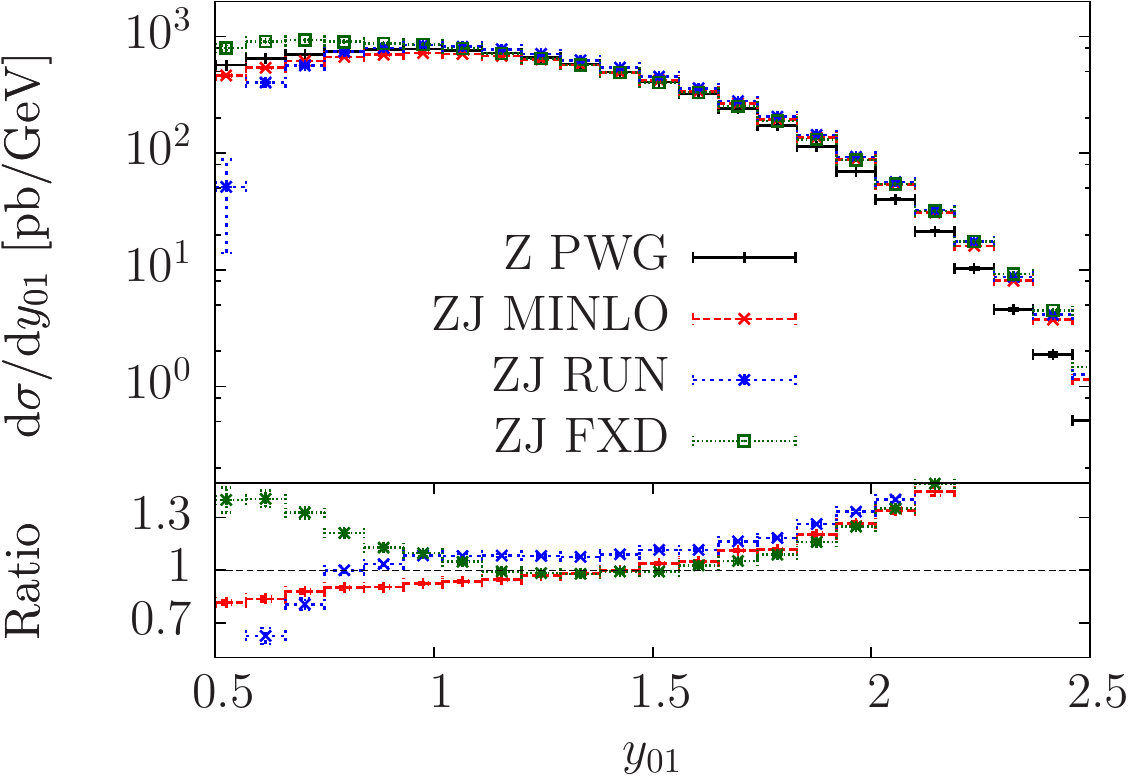} \includegraphics[width=0.495\textwidth]{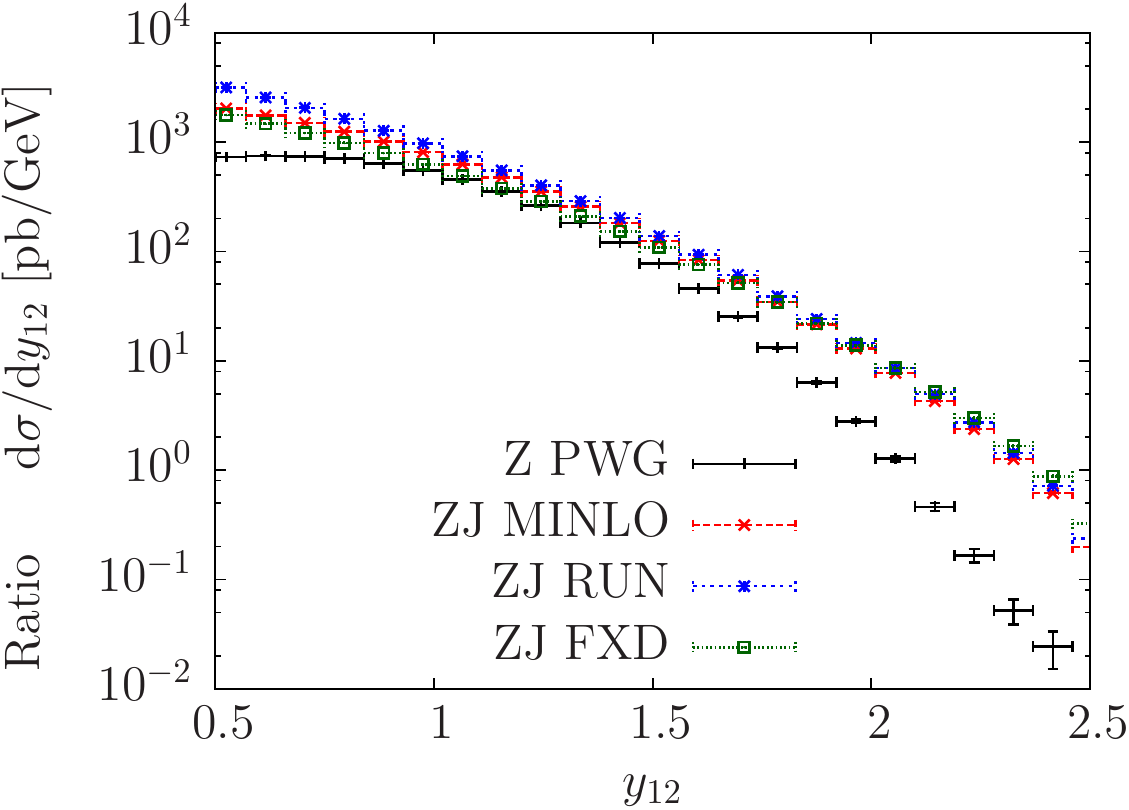}
\par\end{centering}
\caption{
As figure~\ref{fig:ZJ-Z-pt} for the 
$0\rightarrow1$ and $1\rightarrow2$ differential jet rates in inclusive
Z boson production. \label{fig:ZJ-y01} }
\end{figure}

In fig.~\ref{fig:ZJ-y01}
we display the differential jet rates for the $0\to 1$ and $1\to 2$
transitions. In the first case, we see a more realistic
behaviour of the \MINLO{} result with respect to the conventional
NLO predictions, while for the $1\to 2$ transition, as noted previously
in the case of Higgs production, all three methods are unreliable
at small $y_{12}$. This behaviour is of course expected, since Sudakov
resummation of the real radiation is absent in all but the \POWHEG{}+\PYTHIA{}
prediction.

\subsubsection{$Z$ boson production in association with two jets}\label{sec:ZJJ}
\begin{figure}[tbh]
\begin{centering}
\includegraphics[width=0.495\textwidth]{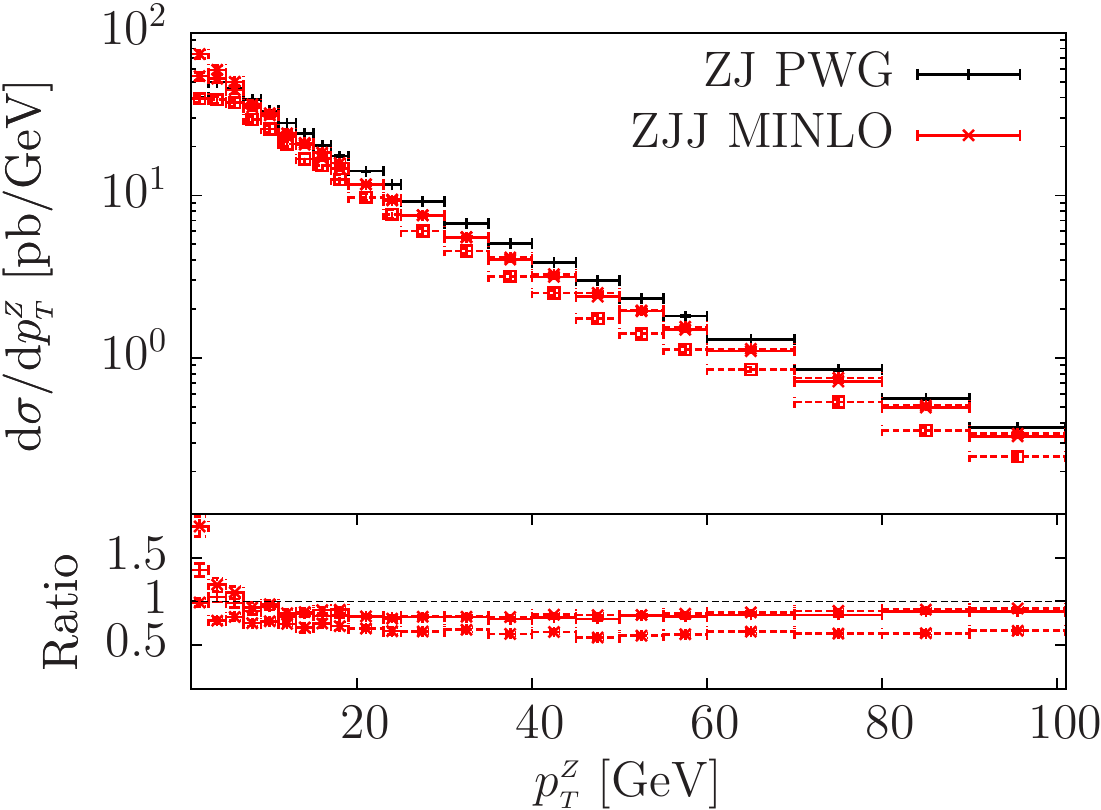}
\includegraphics[width=0.495\textwidth]{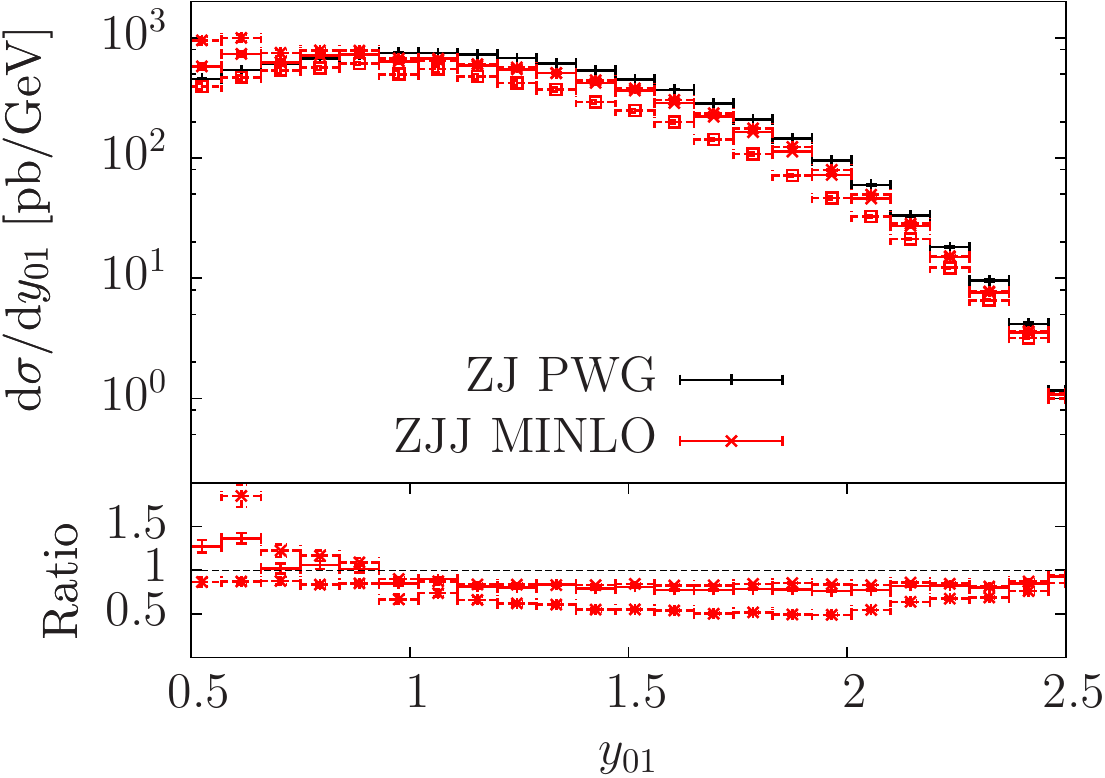} 
\par\end{centering}
\caption{The transverse momentum of the $Z$ boson (left) and the differential
jet rate $y_{01}$ (right), 
computed with the
\POWHEGBOX{} ZJ generator, augmented with the \MINLO{}
procedure and with the \ZJJ{}-\MINLO{} procedure. 
\label{fig:zjj-z-pt}}
\end{figure}

In the case of $Z$ production in association with two jets, one does not
expect a meaningful prediction for the $Z$ transverse momentum and for
the $0\to 1$ differential jet rate from regular NLO calculations with
standard scale choices. As shown in fig.~\ref{fig:zjj-z-pt}
the \MINLO{} result is instead sensible for both distributions.

\begin{figure}[tbh]
\begin{centering}
\includegraphics[width=0.495\textwidth]{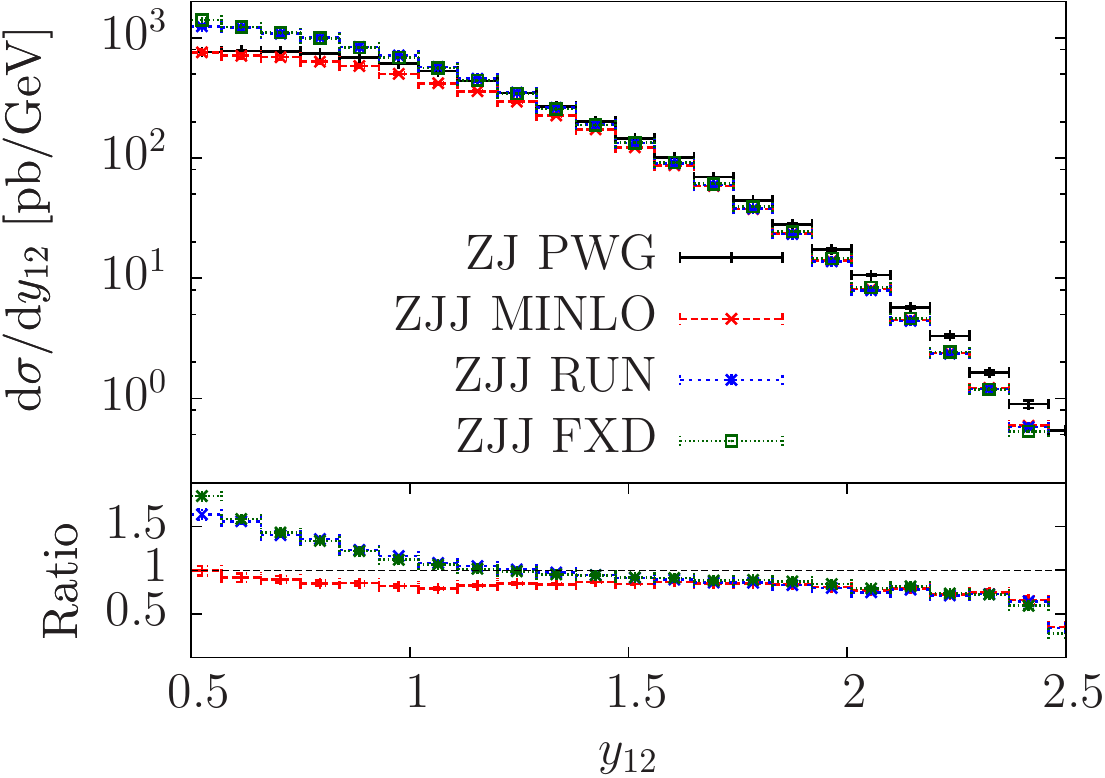}
\includegraphics[width=0.477\textwidth]{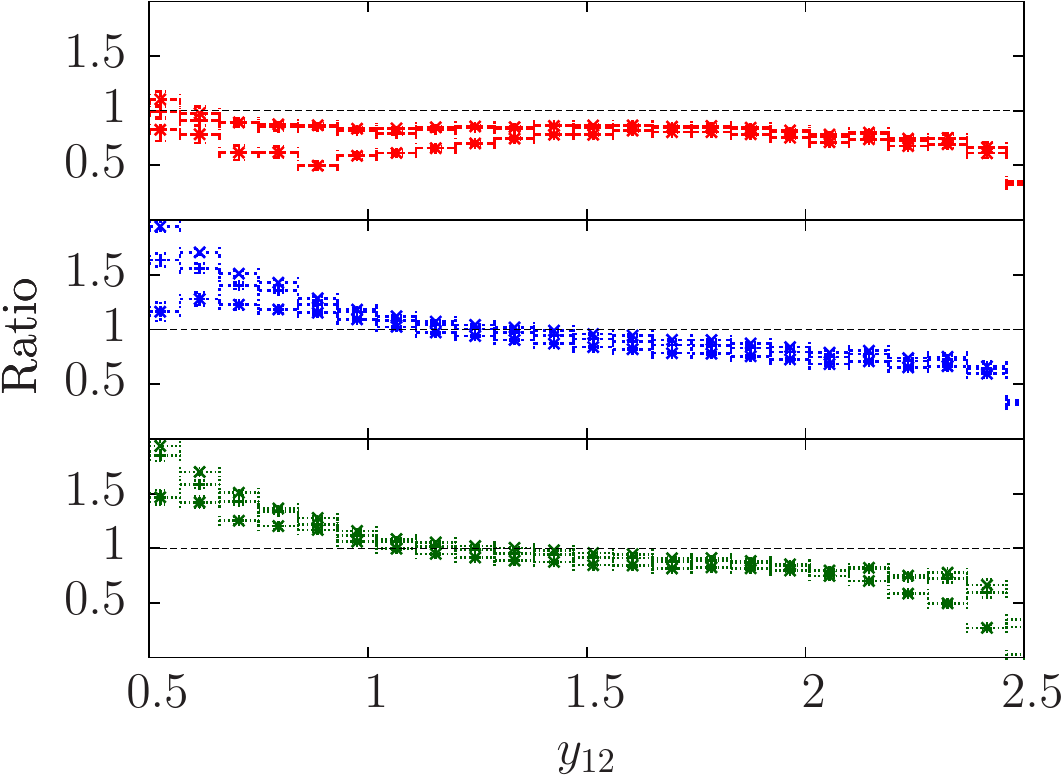} 
\par\end{centering}
\caption{The differential jet rate $y_{12}$, 
computed with the
 \ZJ{} \POWHEGBOX{}
  simulation, augmented by the \MINLO{} procedure (black), 
  the \MINLO{} \ZJJ{} computation (red dashes), and conventional NLO
  \ZJJ{} predictions with $\muF=\muR=\HThat/2$ (blue dots) and
  $\muF=\muR=\mZ$ (fine green dots).
  \label{fig:ZJJ-y12} }
\end{figure}

Following our analysis for Higgs production, in fig.~\ref{fig:ZJJ-y12}
we have superimposed predictions from the \MINLO{} \ZJJ{} and conventional
NLO $Z+2$~jets computations (with different scale choices) on those of the
\ZJ{} \POWHEG{} generator interfaced to \PYTHIA{}, for the $1\to 2$
differential jet rate.
As in the case of Higgs production the Sudakov suppression
effects built into the \MINLO{} prediction are clearly manifest in the
region $y_{12} < 1$, where the regular NLO computations clearly
depart from the fully resummed \ZJ{} \POWHEG{} result.

\begin{figure}[tbh]
\begin{centering}
\includegraphics[width=0.495\textwidth]{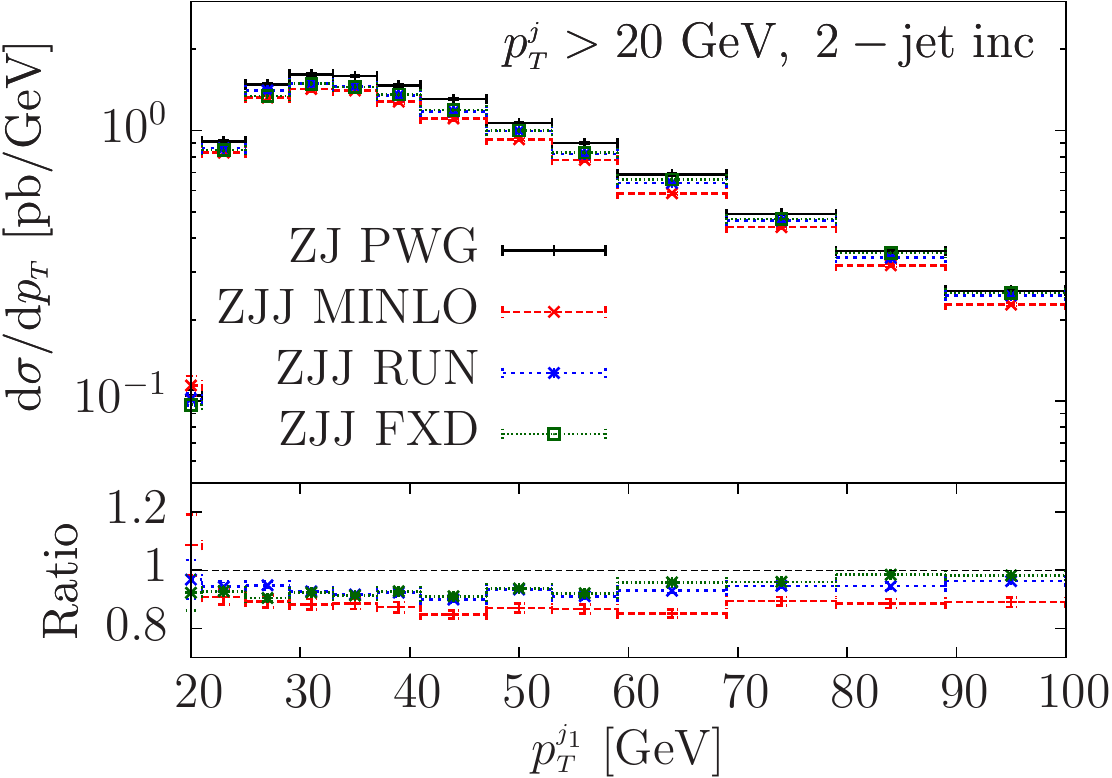}
\includegraphics[width=0.477\textwidth]{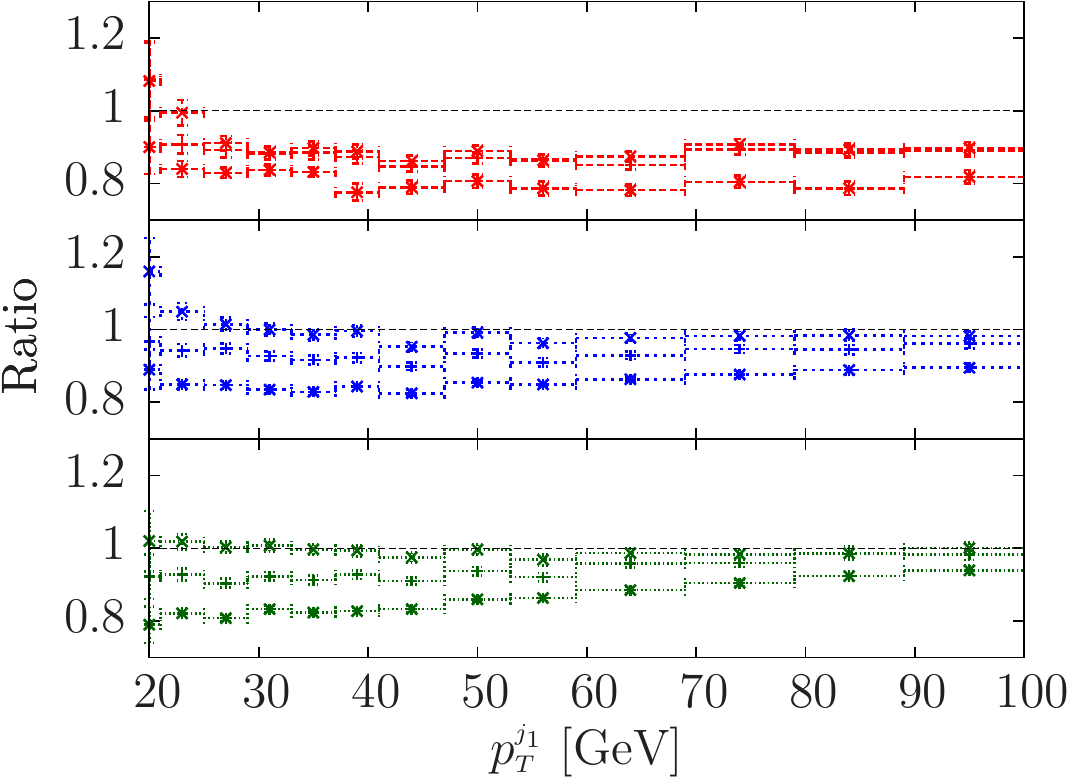} 
\par\end{centering}
\caption{The transverse momentum of the leading jet in events
  comprising the $Z$ boson and at least two jets with $\pT>20$
  GeV\label{fig:ZJJ-Njge2-j1-pt-020}. As in figure \ref{fig:ZJJ-y12},
  on the left we compare the prediction of the \ZJ{} \POWHEG{}
  simulation, augmented by the \MINLO{} procedure (black), to those of
  the \MINLO{} \ZJJ{} computation (red dashes), and conventional NLO
  \ZJJ{} predictions with $\muF=\muR=\HThat/2$ (blue dots) and
  $\muF=\muR=\mZ$ (fine green dots). To the right we show the ratio of
  each of the NLO ZJJ results with respect to the NLO \ZJ{} \POWHEG{}
  simulation, with the band either side of the central values
  indicating the combined renormalization and factorization scale
  uncertainty.}
\end{figure}

\begin{figure}[tbh]
\begin{centering}
\includegraphics[width=0.495\textwidth]{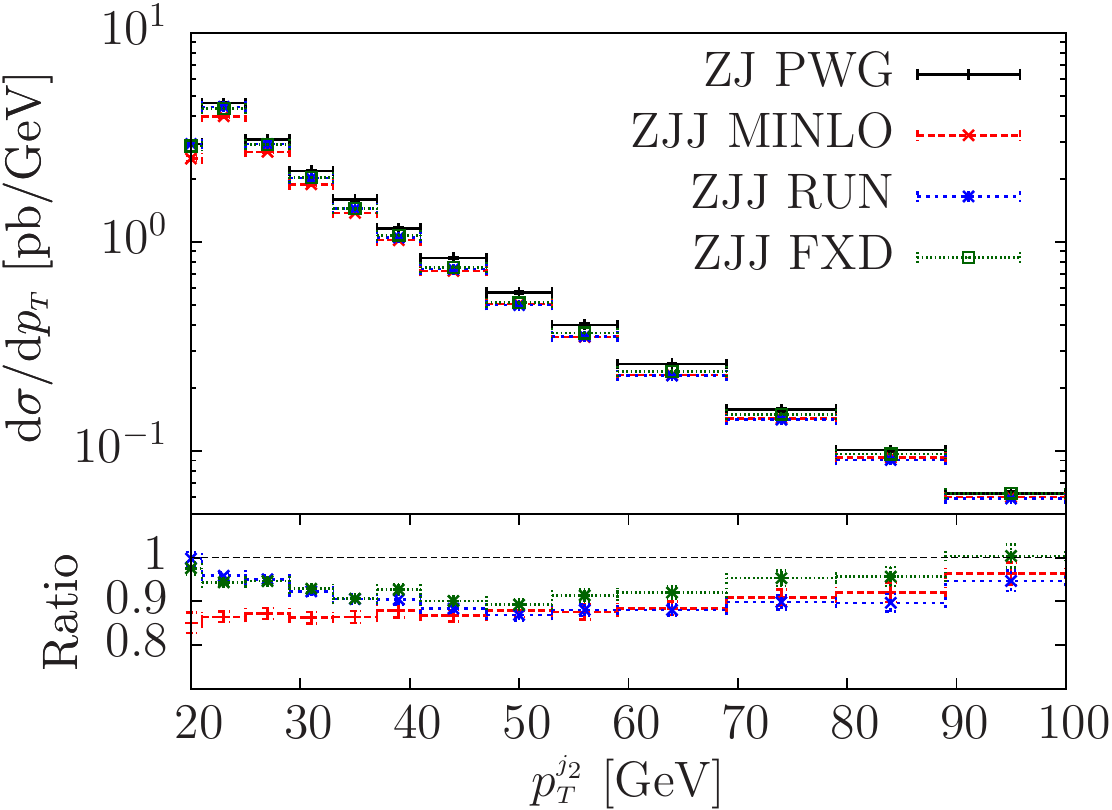}
\includegraphics[width=0.477\textwidth]{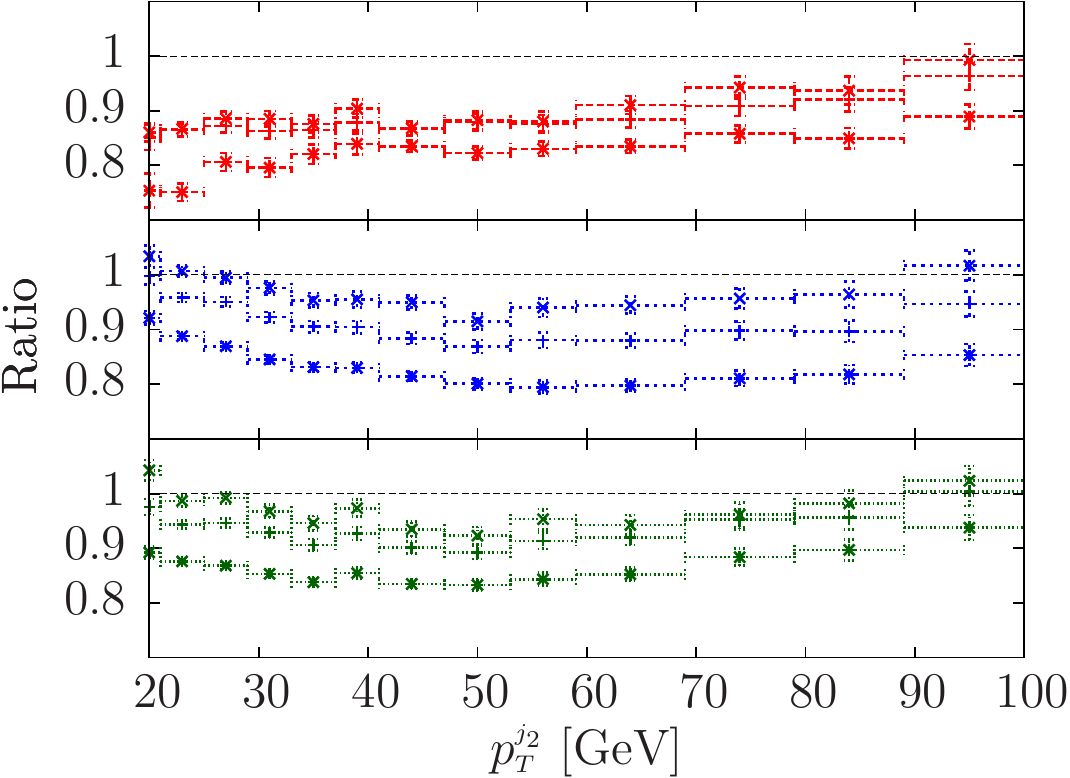} 
\par\end{centering}
\caption{As in figure~\ref{fig:ZJJ-Njge2-j1-pt-020} for the transverse momentum of the next-to-leading jet in
  inclusive $Z$ boson production\label{fig:ZJJ-j2-pt-020}. 
}
\end{figure}

Finally, in figs~\ref{fig:ZJJ-Njge2-j1-pt-020} and \ref{fig:ZJJ-j2-pt-020}
we show the transverse momentum distribution for the first and second
jet, in events with at least two jets. In these figures we see that,
when cuts are imposed to stay away from the Sudakov regions, all
methods are in reasonable agreement.

\section{Conclusions\label{sec:conc}}
In this paper we have formulated a method for the choice of scales and
for the inclusion of Sudakov form factors in NLO calculations of
processes involving jet production. The method proposed is such that
the Born term is evaluated using the CKKW prescription, and the real
and virtual corrections are added in such a way that formal NLO
accuracy is maintained and the good features of the CKKW Born result
are not spoiled.

We have examined the performance of our method (\MINLO{} for
Multi-scale improved NLO) in Higgs and $Z$ boson production in
association with up to two jets.  We have observed the following
properties of the \MINLO{} method as compared to conventional NLO
computations:
\begin{itemize}
\item The \MINLO{} results are well behaved in the Sudakov regions for a large class of distributions,
  where instead the NLO results with standard choice of scale display
  large instabilities and breakdown. Although we do not expect the
  \MINLO{} result to maintain NLO accuracy in the Sudakov regions, we
  clearly see the advantage that when these regions are approached,
  large Sudakov logarithms are exponentiated properly for a large
  class of observables.
\item Away from the Sudakov regions, the \MINLO{} method performs
  similarly to regular NLO computations using standard scale choices
  without Sudakov form factors.
\item In general, NLO results using popular prescriptions for the scale
  choice, like $\HThat/2$, that tend to favour high scales, are in better
  agreement with the \MINLO{} results. We attribute this behaviour as
  being due to the fact that larger scales in general lead to smaller
  cross sections, thus compensating for the lack of genuine Sudakov
  suppression.
\end{itemize}

The method we propose has the further advantage of great simplicity.
All the formulae needed to implement it are given in the present work,
and their \POWHEGBOX{} implementation will soon be made publicly
available.

The advantages of the \MINLO{} method are strictly correlated with the
advantages of the corresponding CKKW procedure. Thus, in the context
of \POWHEG{} simulations including parton showers, augmented with the
\MINLO{} procedure, we expect improved behaviour in Sudakov regions
where more than one invariant becomes small, regardless of the
observable under consideration. This parallels the fact that the CKKW
method used in conjunction with a parton shower generator yields an
improved description of the Sudakov effects for all observables.  If
the \MINLO{} procedure is instead used in the context of a bare NLO
calculation, we expect an improvement for observables that depend upon
the final state pseudo-parton obtained using the $\kT$ clustering
algorithm, at the stage where we have a number of jets equal to the
number $n$ of light partons accompanying the primary process at the
Born level (e.g. $n=1$ for $H+1$~jet and $Z+1$~jet, and $n=2$ for
$H+2$~jets and $Z+2$~jets). For example, distributions constructed out
of the hardest $n$ jets obtained with the inclusive $\kT$-clustering
algorithm satisfy this requirement.

Distributions that do not satisfy the above requirements are not well
described by the \MINLO{} procedure alone. However, we stress again
that the \MINLO{} procedure adopted in conjunction with a NLO+PS
method like \POWHEG{} or \MCatNLO{} should yield an improved
resummation of Sudakov logarithms for all observables.

We thus have been led in this study to conclude that finding an
optimal scale in multi-scale processes requires the inclusion of
Sudakov form factors, whose subleading terms are required to
compensate scale mismatch between nearby vertices.  This leads
formally to improved accuracy for a large class of distributions.
In turn, the inclusion of the Sudakov form factor acts correctly for
all observables only if we match the NLO calculation to a parton
shower algorithm.

As a last point, we remark that many variations can be made on the
method that we have proposed, that do not affect neither the NLO
accuracy, nor the logarithmic resummation. One example is the scale
chosen in the $(N+1)^{\rm th}$ power of $\alphaS$ in the virtual, real
and Sudakov subtraction term. The method that we propose is completely
new, and in the present work we simply present it with a definite
choice among all possible options. We will thus leave the exploration
of all possible alternatives, and eventual refinements of our
prescription, to future work.

\section{Acknowledgments\label{sec:Acknowledgments}}
We thank Gavin Salam and Bryan Webber for useful discussions.  We are
also grateful to Emanuele Re for providing virtual matrix elements
squared for Z+dijet production at one phase space point.

G.Z. is grateful to CERN and to the INFN in Milano-Bicocca for
hospitality while part of this work was carried out.
G.Z.\ is supported by the British Science and Technology Facilities
Council.
G.Z. and P.N. acknowledge the support also of grant
PITN-GA-2010-264564 from the European Commission.

\appendix
\section{NLL improved Sudakov}\label{app:sudak}
We have used the following expression for the Sudakov exponent
\begin{equation}\label{eq:sudexpnll}
\log \Delta_i(Q_0,Q)=-\int_{Q_0}^{Q}\,\frac{{\mathrm d} q^2}{q^2}
\left[\left(a(q^2) A_1^{(i)}+a^2(q^2) A_2^{(i)}\right)\log\frac{Q^2}{q^2}
+a(q^2) B_1^{(i)}\right]\;,
\end{equation}
where
\begin{equation}
A_1^{(q)} = \CF,\quad B_1^{(q)}=-\frac{3}{2}\CF,\quad A_1^{(g)} = \CA,
\quad B_1^{(g)}=-2\pi b_0,
\end{equation}
\begin{equation}
A_2^{(g/q)} = A_1^{(g/q)}\left[\left(\frac{67}{18}-\frac{\pi^2}{6}\right)\CA
-\frac{5}{9} \nF\right]\,,
\end{equation}
and
\begin{equation}
2\pi \,a(q^2)=\alphaS(q^2)=\frac{1}{b_0 \log (q^2/\LambdaMSB^2)}
\left[1-\frac{b_1}{b_0^2}\frac{\log\log( q^2/\LambdaMSB^2) }{\log (q^2/\LambdaMSB^2)}
\right]\,,
\end{equation}
with
\begin{equation}
b_0=\frac{33-2\,\nF}{12\pi},\quad\quad b_1=\frac{153-19\,\nF}{24\pi^2}.
\end{equation}
Equation~(\ref{eq:sudexpnll}) was integrated analytically and the result checked
numerically. We have always used $\nF=5$, and the five flavours expression of $\LambdaMSB$.

\providecommand{\href}[2]{#2}\begingroup\raggedright\endgroup


\begin{thebibliography}{10}

\bibitem{Catani:2001cc}
S.~Catani, F.~Krauss, R.~Kuhn, and B.~R. Webber, {\it {QCD} {M}atrix {E}lements
  + {P}arton {S}howers},  {\em JHEP} {\bf 11} (2001) 063,
  [\href{http://xxx.lanl.gov/abs/hep-ph/0109231}{{\tt hep-ph/0109231}}].

\bibitem{Mangano:2001xp}
M.~L. Mangano, M.~Moretti, and R.~Pittau, {\it {Multijet matrix elements and
  shower evolution in hadronic collisions: $Wb\bar{b}$+n jets as a case
  study}},  {\em Nucl. Phys.} {\bf B632} (2002) 343--362,
  [\href{http://xxx.lanl.gov/abs/hep-ph/0108069}{{\tt hep-ph/0108069}}].

\bibitem{Lonnblad:2001iq}
L.~Lonnblad, {\it {Correcting the colour-dipole cascade model with fixed order
  matrix elements}},  {\em JHEP} {\bf 05} (2002) 046,
  [\href{http://xxx.lanl.gov/abs/hep-ph/0112284}{{\tt hep-ph/0112284}}].

\bibitem{Krauss:2002up}
F.~Krauss, {\it {Matrix elements and parton showers in hadronic interactions}},
   {\em JHEP} {\bf 08} (2002) 015,
  [\href{http://xxx.lanl.gov/abs/hep-ph/0205283}{{\tt hep-ph/0205283}}].

\bibitem{Mrenna:2003if}
S.~Mrenna and P.~Richardson, {\it {Matching matrix elements and parton showers
  with HERWIG and PYTHIA}},  {\em JHEP} {\bf 05} (2004) 040,
  [\href{http://xxx.lanl.gov/abs/hep-ph/0312274}{{\tt hep-ph/0312274}}].

\bibitem{Mangano:2004lu}
M.~Mangano, {\it {Merging multijet matrix elements and shower evolution in
  hadronic collisions}},  {\em
  http://mlm.web.cern.ch/mlm/talks/lund-alpgen.pdf} (2004).

\bibitem{Frixione:2002ik}
S.~Frixione and B.~R. Webber, {\it {Matching NLO QCD Computations and Parton
  Shower Simulations}},  {\em JHEP} {\bf 06} (2002) 029,
  [\href{http://xxx.lanl.gov/abs/hep-ph/0204244}{{\tt hep-ph/0204244}}].

\bibitem{Nason:2004rx}
P.~Nason, {\it {A new method for combining NLO QCD with shower Monte Carlo
  algorithms}},  {\em JHEP} {\bf 11} (2004) 040,
  [\href{http://xxx.lanl.gov/abs/hep-ph/0409146}{{\tt hep-ph/0409146}}].

\bibitem{Frixione:2007vw}
S.~Frixione, P.~Nason, and C.~Oleari, {\it {Matching NLO QCD computations with
  Parton Shower simulations: the POWHEG method}},  {\em JHEP} {\bf 11} (2007)
  070, [\href{http://xxx.lanl.gov/abs/0709.2092}{{\tt arXiv:0709.2092}}].

\bibitem{Ellis:1993tq}
S.~D. Ellis and D.~E. Soper, {\it {Successive combination jet algorithm for
  hadron collisions}},  {\em Phys. Rev.} {\bf D48} (1993) 3160--3166,
  [\href{http://xxx.lanl.gov/abs/hep-ph/9305266}{{\tt hep-ph/9305266}}].

\bibitem{Catani:1993hr}
S.~Catani, Y.~L. Dokshitzer, M.~H. Seymour, and B.~R. Webber, {\it
  {Longitudinally invariant $K_t$ clustering algorithms for hadron hadron
  collisions}},  {\em Nucl. Phys.} {\bf B406} (1993) 187--224.

\bibitem{Rubin:2010xp}
M.~Rubin, G.~P. Salam, and S.~Sapeta, {\it {Giant QCD K-factors beyond NLO}},
  {\em JHEP} {\bf 09} (2010) 084,
  [\href{http://xxx.lanl.gov/abs/1006.2144}{{\tt arXiv:1006.2144}}].

\bibitem{Alioli:2010xd}
S.~Alioli, P.~Nason, C.~Oleari, and E.~Re, {\it {A general framework for
  implementing NLO calculations in shower Monte Carlo programs: the POWHEG
  BOX}},  {\em JHEP} {\bf 06} (2010) 043,
  [\href{http://xxx.lanl.gov/abs/1002.2581}{{\tt arXiv:1002.2581}}].

\bibitem{deFlorian:1999zd}
  D.~de Florian, M.~Grazzini and Z.~Kunszt,
  {\it Higgs production with large transverse momentum in hadronic collisions at next-to-leading order},
  Phys.\ Rev.\ Lett.\  {\bf 82} (1999) 5209
  [hep-ph/9902483].

\bibitem{Ravindran:2002dc}
  V.~Ravindran, J.~Smith and W.~L.~Van Neerven,
  {\it Next-to-leading order QCD corrections to differential distributions of Higgs boson production in hadron hadron collisions},
  Nucl.\ Phys.\ B {\bf 634} (2002) 247
  [hep-ph/0201114].
%
\bibitem{Campbell:2006xx}
  J.~M.~Campbell, R.~K.~Ellis and G.~Zanderighi,
  {\it Next-to-Leading order Higgs + 2 jet production via gluon fusion},
  JHEP {\bf 0610} (2006) 028
  [hep-ph/0608194].

\bibitem{Campbell:2010cz}
  J.~M.~Campbell, R.~K.~Ellis and C.~Williams,
  {\it Hadronic production of a Higgs boson and two jets at next-to-leading order},
  Phys.\ Rev.\ D {\bf 81} (2010) 074023
  [arXiv:1001.4495 [hep-ph]].

\bibitem{Campbell:2012am}
J.~M. Campbell, R.~K. Ellis, R.~Frederix, P.~Nason, C.~Oleari, et~al., {\it
  {NLO Higgs boson production plus one and two jets using the POWHEG BOX,
  MadGraph4 and MCFM}},  \href{http://xxx.lanl.gov/abs/1202.5475}{{\tt
  arXiv:1202.5475}}.

\bibitem{Alioli:2010qp}
S.~Alioli, P.~Nason, C.~Oleari, and E.~Re, {\it {Vector boson plus one jet
  production in POWHEG}},  {\em JHEP} {\bf 01} (2011) 095,
  [\href{http://xxx.lanl.gov/abs/1009.5594}{{\tt arXiv:1009.5594}}].

\bibitem{Re:2012zi}
E.~Re, {\it {NLO corrections merged with parton showers for Z+2 jets production
  using the POWHEG method}},  \href{http://xxx.lanl.gov/abs/1204.5433}{{\tt
  arXiv:1204.5433}}.

\bibitem{Giele:1993dj}
  W.~T.~Giele, E.~W.~N.~Glover and D.~A.~Kosower,
  {\it Higher order corrections to jet cross-sections in hadron colliders},
  Nucl.\ Phys.\ B {\bf 403} (1993) 633
  [hep-ph/9302225].

\bibitem{Bern:1997sc}
Z.~Bern, L.~J. Dixon, and D.~A. Kosower, {\it {One loop amplitudes for e+ e- to
  four partons}},  {\em Nucl.Phys.} {\bf B513} (1998) 3--86,
  [\href{http://xxx.lanl.gov/abs/hep-ph/9708239}{{\tt hep-ph/9708239}}].

\bibitem{Campbell:2002tg}
J.~M. Campbell and R.~K. Ellis, {\it {Next-to-leading order corrections to
  $W^+$ 2 jet and $Z^+$ 2 jet production at hadron colliders}},  {\em
  Phys.Rev.} {\bf D65} (2002) 113007,
  [\href{http://xxx.lanl.gov/abs/hep-ph/0202176}{{\tt hep-ph/0202176}}].

\bibitem{Pumplin:2002vw}
J.~Pumplin, D.~Stump, J.~Huston, H.~Lai, P.~M. Nadolsky, et~al., {\it {New
  generation of parton distributions with uncertainties from global QCD
  analysis}},  {\em JHEP} {\bf 0207} (2002) 012,
  [\href{http://xxx.lanl.gov/abs/hep-ph/0201195}{{\tt hep-ph/0201195}}].

\bibitem{Martin:2009iq}
A.~Martin, W.~Stirling, R.~Thorne, and G.~Watt, {\it {Parton distributions for
  the LHC}},  {\em Eur.Phys.J.} {\bf C63} (2009) 189--285,
  [\href{http://xxx.lanl.gov/abs/0901.0002}{{\tt arXiv:0901.0002}}].

\bibitem{Cacciari:2011ma}
M.~Cacciari, G.~P. Salam, and G.~Soyez, {\it {FastJet user manual}},  {\em
  Eur.Phys.J.} {\bf C72} (2012) 1896,
  [\href{http://xxx.lanl.gov/abs/1111.6097}{{\tt arXiv:1111.6097}}].

\bibitem{Sjostrand:2006za}
T.~Sjostrand, S.~Mrenna, and P.~Z. Skands, {\it {PYTHIA 6.4 Physics and
  Manual}},  {\em JHEP} {\bf 05} (2006) 026,
  [\href{http://xxx.lanl.gov/abs/hep-ph/0603175}{{\tt hep-ph/0603175}}].

\bibitem{Berger:2010zx} 
  C.~F.~Berger, Z.~Bern, L.~J.~Dixon, F.~Febres Cordero, D.~Forde,
  T.~Gleisberg, H.~Ita and D.~A.~Kosower {\it et al.}, {\it Precise
    Predictions for W + 4 Jet Production at the Large Hadron
    Collider}, Phys.\ Rev.\ Lett.\ {\bf 106} (2011) 092001
  [arXiv:1009.2338 [hep-ph]].

\bibitem{GavinPrivate}
G.~P.~Salam, private communication. 

\bibitem{Dittmaier:2012vm}
S.~Dittmaier, S.~Dittmaier, C.~Mariotti, G.~Passarino, R.~Tanaka, et~al., {\it
  {Handbook of LHC Higgs Cross Sections: 2. Differential Distributions}},
  \href{http://xxx.lanl.gov/abs/1201.3084}{{\tt arXiv:1201.3084}}. Report of
  the LHC Higgs Cross Section Working Group.

\bibitem{Bozzi:2008bb}
  G.~Bozzi, S.~Catani, G.~Ferrera, D.~de Florian and M.~Grazzini,
  {\it {Transverse-momentum resummation: A Perturbative study of
     Z production at the Tevatron}}
  Nucl.\ Phys.\ B {\bf 815} (2009) 174
  [\href{http://xxx.lanl.gov/abs/0812.2862}{{\tt arXiv:0812.2862}}].


\end{thebibliography}
\end{document}